\documentclass[prb,twocolumn,preprintnumbers,amsmath,amssymb]{revtex4-1}
\usepackage{graphicx}

\begin{document}
\title{ Photon Berry phases, Instantons, Schrodinger Cats with oscillating parities and crossover from $ U(1) $ to $ Z_2 $ limit in cavity QED systems  }
\author{  Yu Yi-Xiang$^{1,2,3}$, Jinwu Ye $^{1,3}$, W.M. Liu $^{2}$ and CunLin Zhang $^{3}$ }
\affiliation{
   $^{1}$  Department of Physics and Astronomy, Mississippi State  University, P. O. Box 5167, Mississippi State, MS, 39762   \\
   $^{2}$  Beijing National Laboratory for Condensed Matter Physics, Institute of Physics, Chinese Academy of
   Sciences, Beijing 100190, China   \\
   $^{3}$  Key Laboratory of Terahertz Optoelectronics, Ministry of Education, Department of Physics, Capital Normal University, Beijing 100048, China  }
\date{\today }


\begin{abstract}
   The four standard quantum optics models such as
  Rabi, Dicke, Jaynes-Cummings ( JC ) and Tavis-Cummings (TC)  model were proposed by the old generation of great physicists  many decades ago.
   Despite their relative simple forms and many previous theoretical works,
   their solutions at a finite $ N $, especially inside the superradiant regime, remain unknown.
   In this work, we address this outstanding problem by using the $ 1/J $ expansion and exact diagonization
   to study the $ U(1)/Z_2 $ Dicke model at a finite $ N $.
   This model includes the  four standard quantum optics model as its various special limits.
  The $ 1/J $ expansions is complementary to the strong coupling expansion  used by the authors in arXiv:1512.08581 to study the same model
  in its dual $ Z_2/U(1) $ representation.
  We identify 3 regimes of the system's energy levels: the normal, $ U(1) $ and quantum tunneling (QT) regime.
  The system's energy levels are grouped into doublets  which consist of scattering states and Schrodinger Cats
  with even ( e )  and odd ( o ) parities in the $ U(1) $ and quantum tunneling (QT) regime respectively.
  In the QT regime, by the WKB method, we find the emergencies of bound states one by one as the interaction strength increases,
  then investigate a new class of quantum tunneling processes through the instantons between the two bound states in the compact photon phase.
  It is the Berry phase interference effects in the instanton tunneling event which leads to Schrodinger Cats oscillating with even and odd
  parities in both ground and higher energy bound states.
  We map out the energy level evolution from the $ U(1) $ to the QT regime and also discuss some duality relations between the energy levels
  in the two regimes.   We also compute the photon correlation functions, squeezing spectrum, number correlation functions
  in both regimes which can be measured by
  various experimental techniques. The combinations of the results achieved here by $ 1/J $ expansion and those in arXiv:1512.08581
  by strong coupling method lead to rather complete understandings of
  the $ U(1)/Z_2 $ Dicke model at a finite $ N $ and any anisotropy parameter $ \beta $. Experimental realizations and detections
  are presented. Connections with past works and future  perspectives are also discussed.
\end{abstract}

\maketitle

\section{ Introduction }

  Quantum optics is a subject to describe the atom-photon interactions \cite{walls,scully}. The history of quantum optics can be best
  followed by looking at the evolution of quantum optics models to study such interactions.
  In the Rabi model\cite{rabi}, a single mode photon interacts with a two level atom with equal rotating wave
  (RW) and counter rotating wave (CRW) strength.
  To study possible many body effects such as "optical bombs", a single two level atoms in the Rabi model
  was extended to an assembly of $ N $ two level atoms in the Dicke model \cite{dicke}.
  When the coupling strength is well below the transition frequency, the CRW term in the Rabi model is effectively much smaller than that of RW, so
  it was dropped in the Jaynes-Cummings ( JC ) model \cite{jc}. Similar to the generalization from the Rabi to the Dicke model,
  the single two level atom in the JC model
  was extended to an assembly of $ N $ two level atoms in the Tavis-Cummings (TC) model \cite{tc}.

  The importance of the 4 standard quantum optics model at a finite $ N $ in quantum and non-linear optics
  ranks the same as the bosonic or fermionic Hubbard models and
  Heisenberg models in strongly correlated electron systems and the Ising models in Statistical mechanics \cite{aue,sachdev}.
  There have been extensive theoretical investigations on the solutions of the four standard quantum optics model.
  Most of the theoretical works focused on the thermodynamic limit $ N \rightarrow \infty $.
  The TC model was studied at $ N \rightarrow \infty $ in \cite{dicke1,popov,zero,staircase}.
  A normal to a superradiant phase transition was found and the zero mode due to the broken $ U(1) $ symmetry identified in the superradiant phase.
  The Dicke model at $ N \rightarrow \infty $ was investigated in \cite{chaos}.
  A superradiant phase transition with the broken $ Z_2 $ symmetry was found and two gapped modes due to the broken $ Z_2 $ symmetry
  identified in the superradiant phase.  However, there were only very limited works at a finite $ N $.
  The Exact Diagonization (ED) in \cite{chaos} shows that the level statistics in a give parity sector changes from
  the Poissonian distribution in the normal phase to the Wigner-Dyson in the superradiant phase at any finite $ N $.
  For the Dicke model, the ground state photon number at the normal to the superradiant quantum critical point
  QCP  was found \cite{qcphoton,china} to scale as $ \langle n_{ph} \rangle  \sim c N^{1/3} $
  which is a direct consequence of finite size scaling near a QCP with infinite coordination numbers \cite{infinite,extension}.
  There are also formally "exact" Bethe Ansatz-like solution for the integrable TC model at a finite $ N $ \cite{betheu1}.
  Recently, a formal "exact" solution was found even for the non-integrable  Dicke model at $ N=1 $ ( Rabi model )\cite{rabisol}.
  Unfortunately, these "exact" solutions are essentially useless in extracting any physical phenomena \cite{betheu1,rabisol}.

  It is convenient to classify the four well known quantum optics models from a simple symmetry point of view:
  the TC and Dicke model as $ U(1) $ and $ Z_2 $ Dicke model respectively,
  while JC and Rabi model are just as the $ N=1 $ version of the two \cite{berryphase,gold,comment}.
  In fact, as stressed in \cite{gold}, there are also two different representations on all of the 4 models: $ N-$ representation
  with $ N $ independent two level atoms with a large Hilbert space $ 2^N $ and spin $ J=N/2 $ representation with
  a smaller Hilbert space $ N+1 $. The relations between the two representations were  clarified in \cite{gold}.
  The dramatic finite size effects such as the Berry phase effects, Goldstone and Higgs modes on $ U(1) $ Dicke model were thoroughly discussed in both $ N-$ representation in \cite{berryphase} by $ 1/N $ expansion and spin-$ J=N/2 $ representation \cite{gold} by $ 1/J $ representation. Remarkably, we find nearly perfect agreements between the results achieved by $ 1/N $ and $ 1/J $ and the ED studies even when $ N $ gets to its lowest value $ N=1 $.
  The effects of a small CRW term near the $ U(1) $ Dicke model limit was also studied in \cite{gold}.
  In view of the tremendous success of the $1/N $ and $ 1/J $ expansion in studying many strongly correlated electron systems \cite{largen,strongc},
  particularly in the $ U(1) $ Dicke model achieved in \cite{berryphase,gold,comment},
  it is natural to apply them to study the $ Z_2 $ Dicke model as originally planned in \cite{berryphase}.

\begin{figure}
\includegraphics[width=3cm]{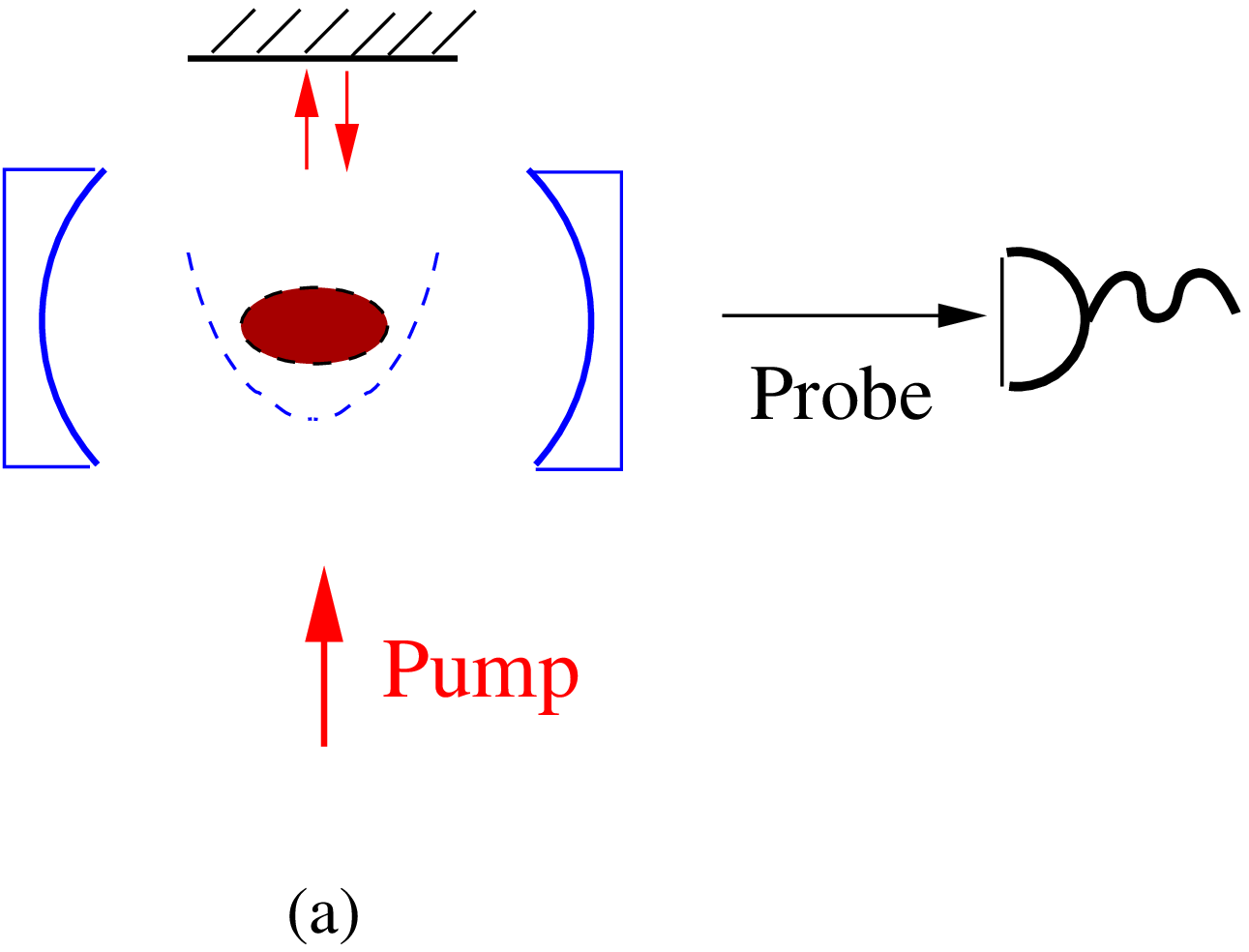}
\hspace{0.5cm}
\includegraphics[width=4 cm]{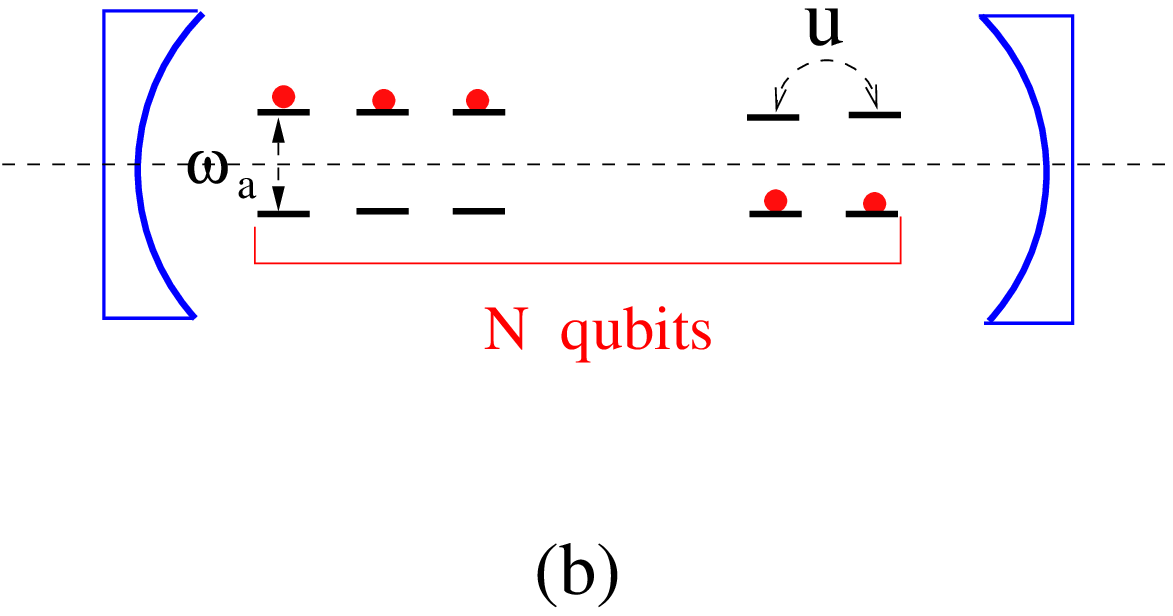}
\caption{ (a) Cold or thermal atoms in a trap embedded in a
cavity in the strong coupling regime with a transverse pumping \cite{orbitalt,orbital,switch}. The probe detects the Fluorescence spectrum of the cavity
leaking photons.  (b) $ N=2\sim 9 $ superconducting qubits are placed on the one or more anti-nodes of a circuit QED resonator
\cite{qubitweak,ultra1,ultra2,qubitstrong,dots}.  }
\label{exp}
\end{figure}

  Due to recent tremendous advances in technologies, the 4 standard quantum optics models were successfully achieved in at least
  two experimental systems (1) with a BEC of $ N \sim 10^5 $ $  ^{87}Rb $ atoms
  inside an ultrahigh-finesse optical cavity \cite{qedbec1,qedbec2,orbitalt,orbital,switch}  and
  (2) superconducting qubits inside a microwave circuit cavity \cite{qubitweak,ultra1,ultra2,qubitstrong}
  or quantum dots inside a semi-conductor microcavity \cite{dots}.
  The superradiant phase in the Dicke model was also realized in system (1) with the help of transverse pumping ( Fig.\ref{exp}a )
  \cite{orbitalt,orbital,switch}. It could also be
  realized "spontaneously " in the system (2) without external pumping \cite{gold}.
  Indeed,  by enhancing the inductive coupling of a flux qubit to a transmission line resonator,
  a remarkable ultra-strong  coupling with individual $ \tilde{g} \sim 0.12 \omega_a $ was realized in a circuit QED system \cite{qubitstrong}.
  However, in such a ultra-strong  coupling regime, the system is described well neither by the TC model nor the Dicke model, but a combination of the two Eqn.\ref{u1z2u1} with unequal RW and CRW strength dubbed as $ U(1)/Z_2 $ Dicke model in \cite{gold}.
  It was also proposed in \cite{gprime1} that in the thermal or cold atom experiments,
  the strengths of $ g $ and $ g^{\prime} $ can be tuned separately by using circularly polarized pump beams in  a ring cavity.
  Indeed, based on the scheme, a recent experiment \cite{expggprime}  realized the $ U(1)/Z_2 $ Dicke model with continuously tunable $ g $ and $ g^{\prime} $.
  As argued in \cite{gold,berryphase}, with only a few $ N=2 \sim 9 $ qubits embedded in system (2),
  the finite size effects may become important and experimentally observable.
  With the recent advances of manipulating a few to a few hundreds of cold atoms inside an optical cavity in system (1) \cite{fewboson,fewfermion},
  the finite size effects may also become important and experimentally observable in near future experiments on system (1).
  As advocated in \cite{gold}, the Hamiltonian Eqn.\ref{u1z2u1} with independent $ g $ and $ g^{\prime} $ is the generic
  Hamiltonian describing various experimental systems  under the two atomic levels and a single photon mode approximation.
  In \cite{gold}, by the $ 1/J $ expansion, we focused on the $ U(1)/Z_2 $ Dicke model Eqn.\ref{u1z2u1} near the $ U(1) $ limit ( namely, with a small
  anisotropy parameter $  g^{\prime}/g=\beta \ll 1 $ and not too far from the critical strength $ g_c $ ) at any finite $ N $.
  In a very recent preprint \cite{strongED}, by the strong coupling expansion and the ED,
  the authors studied the  $ U(1)/Z_2 $ Dicke model in its dual presentation  starting from from the $ Z_2 $ limit $ \beta=1 $.
  Here, by the $ 1/J $ expansion and ED, we will study  the $ U(1)/Z_2 $ Dicke model Eqn.\ref{u1z2u1}
  starting from from the $ U(1) $ limit $ \beta=0 $ which is complementary to the strong coupling expansion in \cite{strongED}.
  The combinations of both approaches will lead to rather complete understandings of the $ U(1)/Z_2 $ Dicke model Eqn.\ref{u1z2u1}
  in the full range of $ 0< \beta < 1 $.

\begin{figure}
\includegraphics[width=7cm]{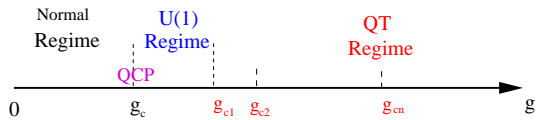}
\caption{ The three regimes of the $ U(1)/Z_2 $ Dicke model at a finite $ N $  as the coupling $ g $ increases
at a fixed anisotropy parameter $ 0<  g^{\prime}/g=\beta < 1 $.
The normal regime, the $ U(1) $ regime and the quantum tunneling (QT) regime.
The quantum critical point ( QCP ) at $ N=\infty $ is at $ g_c= \sqrt{ \omega_a \omega_b}/( 1 + \beta )  $.
The superradiant regime at $ N=\infty $ splits into the $ U(1) $ regime and the QT regime at a finite $ N $.
The dynamic variables in the three regimes are: $ a, b $ in the normal regime, $ \delta \rho_a, e^{i \theta_a} $ in the
$ U(1) $ regime and $ \delta \rho_a, \theta_a, \tau_z $ in the QT regime.
As shown in the text, there exists some duality relations between the $ U(1) $ regime and the QT regime.
The strong coupling expansion in \cite{strongED} works very well in the QT regime and also $ U(1) $ regime not too close to the QCP.
The $ U(1) $ regime may disappear if $ \beta $  gets too close to 1.  }
\label{crossover}
\end{figure}

  In this paper, we study novel quantum phenomena in the $ U(1)/Z_2 $ Dicke model Eqn.\ref{u1z2u1} in its spin $ J=N/2 $ representation
  at a finite $ N $, any interaction strength $ g $ and anisotropy parameter $ 0 < \beta  < 1 $  by
  the $ 1/J $ expansion \cite{gold} and the ED \cite{chaos,gold}.
  As a fixed $ \beta $, as the $ g $  increases, we identify 3 crossover regimes: the normal, $ U(1) $ and
  the quantum tunneling (QT) regime Fig.\ref{crossover}.
  The super-radiant regime at $ N=\infty $ splits into the two regimes at a finite $ N $: the $ U(1) $ and quantum tunneling (QT) regime.
  In the $ U(1) $ regime, we perform a (non-)degenerate perturbation to evaluate the energy spectrum.
  It is the Berry phase which leads to the level crossings between the even and odd parity, therefore the alternating parities on the
  ground state and excited states.
  In the QT regime, by the WKB method, we find the emergencies of bound states one by one as the interaction strength increases,
  then investigate a new class of quantum tunneling processes through the instantons between the two bound states in the compact photon phase.
  It is the Berry phase interference effects in the instanton tunneling event which leads to Schrodinger Cats oscillating with even and odd
  parities in both ground and higher energy bound states.
  We map out the energy level evolution from the $ U(1) $ to the QT regime.
  In the $ U(1) $ regime, the doublets consist of scattering states organized as $ (e,o),(o,e).....$
  ( or $ (o,e),(e,o).....$ ) pattern. While in the QT regime,
  the doublets consist of bound states ( or Schrodinger Cats ) organized as $ (e,o),(e,o).....$  ( or $ (o,e),(o,e).....$ ) pattern.
  There are also some sort of duality relations  between the two regimes in both Hamiltonian and spectrum.
  We  compute the photon correlation functions, squeezing spectrum and number correlation functions in both regimes which can be
  detected by Fluorescence spectrum, phase sensitive homodyne detection and Hanbury-Brown-Twiss (HBT) type of experiments respectively \cite{exciton}.
  When comparing the results with those achieved from the strong coupling expansion in \cite{strongED},
  we find nearly perfect agreements among the $1/J $ expansion, the strong coupling expansion and the ED not only in the QT regime, but also
  in the $ U(1) $ regime not too close to the QCP at $ N=\infty $.
  The combination of the three methods lead to rather complete physical pictures
  in the whole crossover regimes from the $ U(1) $ Dicke to the $ Z_2 $  Dicke model in Fig.\ref{crossover}.
  Experimental realizations, especially the preparations and detections of the Schrodinger Cats
  in both experimental systems are discussed. Connections with the previous works are speculated and future  perspectives are outlined.

\section{  $ 1/J $ expansion in the super-radiant phase. }
   The $ U(1)/Z_2 $ Dicke model\cite{gold,comment} is described by:
\begin{eqnarray}
  H_{U(1)/Z_{2}} &  = &  \omega_a a^{\dagger} a + \omega_b  J_{z}
  + \frac{g}{\sqrt{2J}} (  a^{\dagger} J_{-}+ a  J_{+}  )
      \nonumber   \\
  &  + &   \frac{g^{\prime}}{\sqrt{2J}} (  a^{\dagger} J_{+}+ a J_{-} )
\label{u1z2u1}
\end{eqnarray}
   where the $ \omega_a, \omega_b $ are the cavity photon frequency and the energy difference of the two atomic levels  respectively,
   the  $  g= \sqrt{N} \tilde{g}, N=2J $ is the collective photon-atom rotating wave (RW) coupling.
   The $  g^{\prime}= \sqrt{N} \tilde{g}^{\prime} $ is the counter-rotating wave (CRW) term.
   We fix their ratio to be $ 0< g^{\prime}/g = \beta < 1  $.
   If $ \beta=0 $,  Eqn.\ref{u1z2u1} reduces to the $ U(1) $ Dicke model \cite{berryphase,gold} with the $ U(1) $ symmetry
   $ a \rightarrow  a  e^{ i \theta}, \sigma^{-} \rightarrow \sigma^{-} e^{ i \theta} $ leading to the conserved quantity
   $ P=  a^{\dagger} a + J_z $.  The CRW $ g^{\prime} $ term breaks the $ U(1) $ to the $ Z_2 $ symmetry
   $ a \rightarrow -a , \sigma^{-} \rightarrow -\sigma^{-} $ with the conserved parity operator  $ \Pi= e^{ i \pi ( a^{\dagger} a + J_z ) } $.
   If $ \beta=1  $, it becomes the $ Z_2 $ Dicke model \cite{chaos,qcphoton}.

   Following \cite{gold}, inside the super-radiant phase,  it is convenient to write
   both the photon and atom in the polar coordinates $ a= \sqrt{
   \lambda^{2}_{a} + \delta \rho_a } e^{ i \theta_a}, b= \sqrt{
   \lambda^{2}_{b} + \delta \rho_b } e^{ i \theta_b} $. When
   performing the controlled $ 1/J $ expansion, we keep the terms to the order
   of $ \sim j, \sim 1 $ and $ \sim 1/j $, but drop orders of $ 1/j^{2} $ or higher.
   We first minimize the ground state energy at the order $ j $ and
   found the saddle point values of $ \lambda_a $ and $ \lambda_b $:
\begin{equation}
 \lambda_a  =  \frac{ g + g^{\prime} }{ \omega_a } \sqrt{ \frac{j}{2} ( 1 - \mu^{2} )
  },~~~~ \lambda_b= \sqrt{ j(1-\mu) }
\label{meanz2}
\end{equation}
   where $  \mu = \omega_a \omega_b/(g + g^{\prime} )^{2} $. In the superradiant phase
   $ g+g^{\prime} > g_c= \sqrt{\omega_a \omega_b} $. In the
   normal phase $ g+g^{\prime} < g_c $, one gets back to the trivial solution $
   \lambda_a=\lambda_b = 0 $.

   Observe that (1) in the superradiant phase $ g( 1+ \beta ) > g_c $, $ \lambda^{2}_a \sim
   \lambda^{2}_b \sim j $, (2) it is  convenient to introduce the $ \pm $ modes:
   $ \theta_{\pm}= (\theta_a \pm \theta_b)/2, \delta \rho_{\pm}= \delta \rho_a \pm \delta \rho_b,
   \lambda^{2}_{\pm}= \lambda^{2}_a \pm  \lambda^{2}_b $.
   (3) Defining the Berry phase in the $ + $ sector \cite{berryphase} as $ \lambda^{2}_{+}= P + \alpha $ where $ P=1,2,\cdots $ is the closest integer  to the $ \lambda^{2}_{+} $, so $ -1/2 < \alpha < 1/2 $.
   Due to the large gap in the $ \theta_{-} $, it is justified to drop the Berry phase in the $ - $ sector.
   (4) after shifting $ \theta_{\pm} \rightarrow \theta_{\pm} + \pi/2 $,  we reach the Hamiltonian to the order of $ 1/j $:
\begin{eqnarray}
 {\cal H}[ \delta \rho_{\pm}, \theta_{\pm} ] & = &
 \frac{D}{2} (\delta \rho_{+} - \alpha )^{2} + D_{-} [\delta \rho_{-} + \gamma  ( \delta \rho_{+}-\alpha )]^{2}
               \nonumber    \\
   & + &  4 \omega_{a} \lambda^{2}_a [ \frac{ 1 }{ 1+ \beta } \sin^{2} \theta_{-}
     +  \frac{ \beta }{ 1+ \beta } \sin^{2} \theta_{+} ]
\label{pmu1z2h}
\end{eqnarray}
   where $ D=  \frac{ 2 \omega_a (g+ g^{\prime} )^{2} }{  E^{2}_{H} N }  $ is the phase diffusion constant in the $ + $ sector,
   $  D_{-}= E^{2}_{H}/16 \lambda^{2}_{a} \omega_a $ with $  E^{2}_{H}= ( \omega_a+\omega_b)^{2} + 4 ( g+ g^{\prime} ) ^2 \lambda^{2}_{a}/N $.
   The $ \gamma= \frac{ \omega^{2}_{a} }{ E^{2}_{H} } ( 1- \frac{ ( g + g^{\prime} )^{4}}{ \omega^{4}_{a} } ) $ is the coupling between the $ + $ and $ - $ sector.

    Note that the large $ J $ expansion condition is $ \lambda^{2}_a \gg 1 $.
    In the $ j \rightarrow \infty $ limit, it holds for any $ g > g_c $, but leads to a constraint on $ g $  at a finite $ j $.
    In the superradiant regime, one can simply set $ \sin^{2} \theta_{\pm} \sim \theta^2_{\pm} $ in
    Eqn.\ref{pmu1z2h} which becomes a quadratic theory. It can be easily diagonalized and  lead to one low energy gapped pseudo-Goldstone mode
    and a high energy gapped optical mode. Setting $ \beta=1 $ recovers the results for the $ Z_2 $ Dicke model in the superradiant phase
    at $ N=\infty $ in \cite{chaos}.

      If one neglects the quantum fluctuations of the $ \theta_{-} $ mode, namely, by setting $ \theta_{-} $ at its classical value
   $ \theta_{-}=0 $, Eqn.\ref{pmu1z2h} is simplified to:
\begin{equation}
  {\cal H}_{+} [  \delta \rho_{+}, \theta_{+} ]   = \frac{ D }{2}  ( \delta \rho_{+} -\alpha  )^{2}
    +  2 \omega_{a} \lambda^{2}_a \frac{ 2 \beta }{ 1+ \beta }  \sin^2 \theta_{+}
\label{phaselowh}
\end{equation}
    where $ [ \theta_{+}, \delta \rho_{+} ]= i \hbar $.
  In the superradiant limit $  4 \lambda^{2}_a \frac{ \beta }{ 1+ \beta } \gg 1  $, one can identify the approximate atomic mode:
\begin{equation}
  \omega^{2}_{-0}= 4 \omega_{a} \lambda^{2}_a \frac{ 2 \beta }{ 1+ \beta } D= \frac{4}{ E^{2}_{H} } \frac{ \beta }{ 1+ \beta }
  [ ( g+  g^{\prime})^{4}- g^{4}_{c} ]
\label{atomicm}
\end{equation}
   which is nothing but the pseudo-Goldstone mode due to the CRW $ g^{\prime} $ term.

   Note that for small $ \beta < 1 $, the condition to reach the superradiant regime $  4 \lambda^{2}_a \frac{ \beta }{ 1+ \beta } \gg 1  $
   is more stringent than the large $ J $ expansion condition $ \lambda^{2}_a \gg 1 $.

   By neglecting the quantum fluctuations of $ \theta_{-} $,
   the high energy optical mode in the $ \theta_{-} $ sector can not be seen anymore in Eqn.\ref{phaselowh}.
   The approximation may not give very precise numbers to physical quantities, but do lead to
   correct qualitative physical picture, especially the topological effects due to the Berry phase in all the physical quantities.


\begin{figure}
\includegraphics[width=8.5cm]{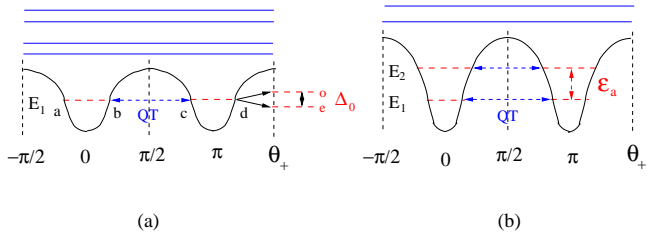}
\caption{ The bound states and the quantum tunneling processes in the QT regime in Fig.\ref{crossover}.
 The atomic energy $ \epsilon_a $ and the tunneling energy $ \Delta_0 $ are shown.
 But the higher optical energy $  \epsilon_o $ is not shown.
(a)  At $ g=g_{c1} > g_c=\sqrt{ \omega_a \omega_b }/( 1 + \beta ) $, the double well potential in the $ \theta_{+} $
sector in Eqn.\ref{phaselowh} holds just one bound state with energy $ E_1 $ denoted by a
red dashed line. The blue dashed line shows
the quantum tunneling between the two bound states which leads to the
splitting $ \Delta_0 $ listed in Eqn.\ref{splitting}. The blue solid
lines denote  scattering states shown in Fig.\ref{instanton}a,\ref{levelevolution}b (b) As the $ g $ increases
further to $ g=g_{c2} > g_{c1}$, the lowest two scattering states in
(a) also become the second bound state with energy $ E_2=E_1+
\epsilon_a $, the third bound state shows up at $ g=g_{c3} > g_{c2}
> g_{c1}  $ and so on ( see also Fig.\ref{levelevolution} ). The excitation energy is the "atomic" energy $ \epsilon_a $.
As explained below Eqn.\ref{evenodd0}, the ground state could also be odd parity depending on $ (-1)^{P} $.  }
\label{bound}
\end{figure}

\section{  $ U(1) $ regime and the formation of consecutive bound states in quantum tunneling  regime. }
   As the potential in the $ \theta_{+} $ sector in Eqn.\ref{phaselowh} gets deeper and deeper, there are
   consecutive bound states formations at $  g_c < g_{c1} < g_{c2} \cdots  $
   leading to the "atomic" energy scale $ \epsilon_{a} $. The QT regime in Fig.\ref{crossover} is
   signatured by the first appearance of the bound state after which there are
   consecutive appearances of more bound states at higher energies ( Fig.\ref{bound}a,b  ).
   The regime $ g_c < g < g_{c1} $ is the $ U(1) $ regime in Fig.\ref{crossover}.

    One can calculate all these $ g_c < g_{c1} < g_{c2} \cdots  $  by using the Bohr-Sommerfeld quantization condition
    for a smooth potential $
    \int^{b}_{a}  p d \theta = ( n + 1/2 ) \pi \hbar,~n=0,1,2,\cdots $
    where  $ p=\sqrt{ 2m( E_{n+1} - V(\theta))} $  and the $ E_{n+1}=E_1, E_2,....... $ is the $ (n+1)-th $ bound state energy in Fig.\ref{bound}a,b.
    From Eqn.\ref{phaselowh}, we can see the $ m = \frac{1}{ D }, V(\theta)=  \omega_{a} \lambda^{2}_a \frac{ 2 \beta }{ 1+ \beta }( 1-\cos 2 \theta_{+}) $ and
    the $ a $ and $ b $ are the two end points shown in Fig.\ref{bound}a.  We find that the bound states emerge at
\begin{equation}
   \frac{ \omega_{-0} }{ D} = ( n + 1/2 ) \frac{\pi }{2}  \hbar,~~~ n=0,1,2,\cdots
\label{boundn}
\end{equation}

    In Eqn.\ref{boundn}, setting $ n=0 $, one can see
    when  $   \frac{ \omega_{-0} }{ D} < \frac{ \pi }{4} \hbar $, there is no bound state.
    This is the $ U(1) $ regime in Fig.\ref{crossover}.
    Substituting the expression for the phase diffusion constant $ D $ and the atomic mode  $ \omega_{-0} $ in Eqn.\ref{atomicm}
    leads to the condition for the $ U(1) $ regime:
\begin{equation}
    4 \lambda^2_{a} \sqrt{ \frac{ \beta }{ 1+ \beta }} < \frac{ \pi }{4}
\label{u1cross}
\end{equation}
    Note that for large $ J $ expansion to apply, one only need to require $  \lambda^2_{a} > 1 $,
    So for sufficiently small $ g^{\prime}/g=\beta $, there is an appreciable $ U(1) $ regime $ g_c < g < g_{c1} $ before
    the quantum tunneling (QT) regime in Fig.\ref{crossover}.

\begin{widetext}

\begin{figure}
\includegraphics[width=15cm]{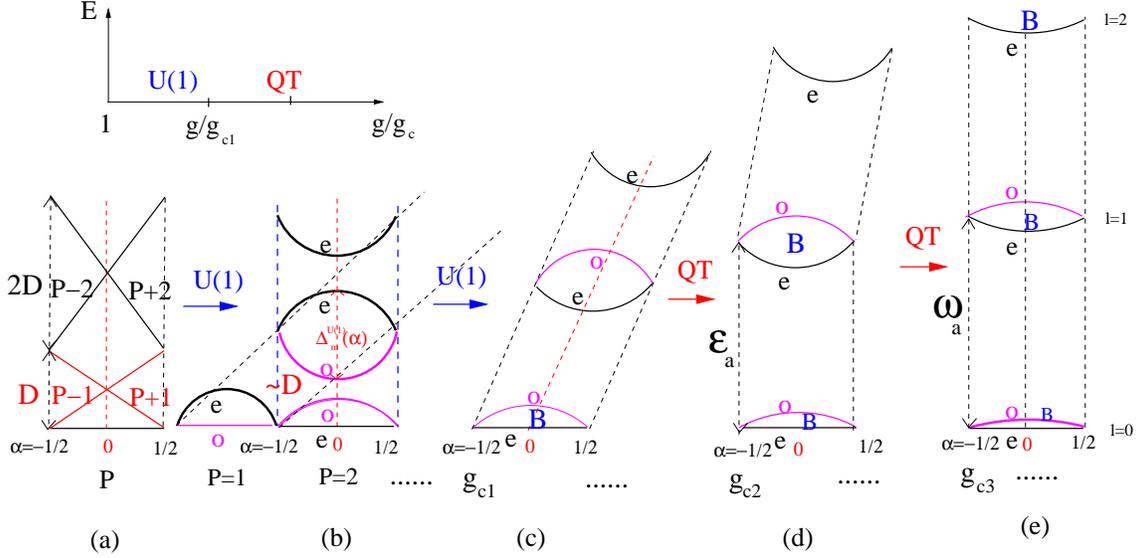}
\caption{ The atomic energy level evolution from the $ U(1) $ regime to the QT regime as $ g $ changes at a fixed $ \beta $.
 The ground state energy has been subtracted.  $ D $ is the phase diffusion constant.
 In the $ U(1) $ regime in (a) and (b), there are $ P $ doublets along the vertical  ( blue dashed ) lines.
 (a) The energy spectrum at the $ U(1) $ limit $ \beta=0 $ \cite{gold}.
 (b) The energy level repulsions between the same parity at $ \alpha=0 $ and level crossings at $ \alpha=\pm 1/2 $
 between the opposite parities in the $ U(1) $ regime. The maximum splitting at  $ \alpha=0 $ and the
 gap at the edge $ \alpha=\pm 1/2 $ are given by Eqn.\ref{deltau1m} and \ref{u1gap} respectively.
 All the states are scattering states. If the total excitation $ P $ is even,
 along the vertical  ( blue dashed ) lines, the doublets are organized as $ (e,o),(o,e),(e,o),(o,e),\cdots $ ( see also Fig.\ref{photonsplitting}a ).
 If $ P $ is odd, one need flip all the parities.  See Fig.\ref{leveledn2}a,\ref{zeros159}a at $ N=2, \beta=0.1 $.
 If we follow the states with $ P, P+1, P+2.... $ at $ l=0,1,2..$ respectively, then the doublets are organized as
 $ (e,o),(e,o),\cdots $, there are relatively large shifts of zeros to the right delineated by the black dashed line.
 Only the splittings at $ m=\pm1, \pm 2 $ at $ P=1,2 $ are shown here.
 See Fig.\ref{leveledn2}b,\ref{zeros159}b at $ N=2, \beta=0.5 $.
 (c) The doublet at $ l=0 $ becomes the first bound state ( Schrodinger cat ) denoted as $ B $ at $ g=g_{c1} $,
 while the states at $ l=1...$ remain scattering states as shown in Fig.\ref{bound}a,\ref{instanton}a.
 (d) The doublet at $ l=1 $ becomes the second bound state at $ g=g_{c2} $, the two bound states at $ l=0,1 $
     are connected by nearly straight boundaries at $ \alpha=-1/2, 0, 1/2$. The states at $  l=2,....$ remain scattering states
     as shown in Fig.\ref{bound}b.  See Fig.\ref{leveledn2}c,\ref{zeros159}c at $ N=2, \beta=0.9 $.
 (e) Finally, more states become bound states at $ g=g_{c3},\cdots $ where the energy level pattern becomes
     $ (e,o),(e,o),\cdots $ in the QT regime ( see also Fig.\ref{photonsplitting}b ). If $ P $ is odd, one need also flip all the parities.
     There is a shift by exact one period from (b) to (e). Intuitively, a leaning tower with infinite stories gradually becomes straight as the system evolves from the $ U(1) $ regime to the QT regime.
     The maximum splitting at $ \alpha=0 $ are given by Eqn.\ref{deltau1},\ref{deltau1m} for scattering states, Eqn.\ref{splitting},\ref{splittingn},
     for bound states ( Schrodinger cats ) and decrease from (b) to (e).
     The edge gap at $ \alpha=\pm 1/2 $ are given by Eqn.\ref{u1gap} and
     the atomic energy $ \epsilon_a $ in Fig.\ref{bound}b for scattering states and  bound states respectively and  increases from (b) to (e),
     but starts to become flat  since the second bound state forms at $ l=1 $ in (d).
     The maximum gap at $ \alpha=0 $ is comparable to the edge gap at  $ \alpha=\pm 1/2 $ in the $ U(1) $ regime (b),
     but much smaller in the QT regime where $ \Delta_l \ll  \omega_a $  (d).
     This evolution from (b) to (e) is precisely observed in the ED in Fig.\ref{leveledn2} and fine structures in all the doublets at
     $ l=0,1,2 $ in Fig.\ref{zeros159}.
     The higher energy optical modes are not shown.
     As said in the caption of Fig.\ref{crossover}, the $ U(1) $ regime disappears  if $ \beta $
     gets too close to $ 1 $ as shown in Fig.\ref{leveledn2}c,\ref{zeros159}c. }
\label{levelevolution}
\end{figure}

\end{widetext}

   In this $ U(1) $ regime, the second term in Eqn.\ref{phaselowh} breaks the $ U(1) $ symmetry to $ Z_2 $ symmetry,
   the Goldstone mode at $ N=\infty $ simply becomes a pseudo-Goldstone mode \cite{berryphase,gold,comment}.
   But its effects at a finite $ N $ is much more delicate to analyze.
   One can treat the second term in Eqn.\ref{phaselowh} perturbatively either by
   a non-degenerate at $ \alpha \neq 0 $ or degenerate perturbation expansion at $ \alpha = 0 $.
   The total excitation $ P $ is not conserved anymore and is replaced by
   the conserved parity  $ \Pi=(-1)^P $, the energy levels are only grouped into even and
   odd parities in Fig.\ref{levelevolution}.
   As shown in \cite{gold}, at a given sector  $ P $, $ m=0,\pm 1,\cdots, \pm P $ at $ \alpha=0 $,
   a first order degenerate perturbation at $ m=\pm 1 $ leads to the maximum splitting at  $ \alpha=0 $ in Fig.\ref{levelevolution}b:
\begin{equation}
   \Delta_{m=\pm 1, U(1)}(\alpha=0 ) =\frac{D}{2}- \frac{ \omega_a \lambda^2_{a} }{2} \frac{ 2 \beta }{ 1 + \beta }
\label{deltau1}
\end{equation}

   Using Eqn.\ref{atomicm}, one can rewrite $ \Delta= \frac{D}{2}[1-\frac{1}{2}( \frac{\omega_{-0}}{D} )^2 ] > 0 $.
   For general $ \pm m $, one needs $ m-th $ order ( with the constraint $ |m | \leq P $ in a given sector $ P $ )
   degenerate perturbation calculation to find the
   the maximum splitting $ \Delta_{ m, U(1)} $ at $ \alpha=0 $ between the $ m $ and $ m+ 1 $ crossing in Fig.\ref{levelevolution}b:
\begin{equation}
   \Delta_{m, U(1)}(\alpha=0 )= D ( m + \frac{1}{2} )- R_{m+1}-R_{m}
\label{deltau1m}
\end{equation}
    where  $  R_m \sim ( \frac{ \omega_a \lambda^2_{a} }{2} \frac{ 2 \beta }{ 1 + \beta })^{ m }, m=0, 1,....P; R_0=0 $ is the gap opening
    at the $ m+1 $ crossing in the $ U(1) $ limit Fig.\ref{levelevolution}a. Setting $ m=0 $ recovers Eqn.\ref{deltau1}.

    Note that the degenerate pair $ (m, -m-1) $ at the edge at $ \alpha=\pm 1/2 $ has different parity, so will not be mixed in any
    order of perturbations \cite{gold}.  It is easy to compute the edge gap at $ \alpha=\pm 1/2 $ by a non-degenerate perturbation.
    Obviously, the second term in Eqn.\ref{phaselowh} connects only $ \Delta m = \pm 2 $, so the first order perturbation
    vanishes, one need to get to at least second order non-degenerate perturbation \cite{gold}. There could also be a slight shift
    in the crossing point between the two opposite parities at $ (m, -m-1) $
\begin{equation}
   \Delta_{m, U(1)}(\alpha=\pm 1/2 )=  D ( m + 1 )- S(m)
\label{u1gap}
\end{equation}
   where $  S(m) \sim ( \frac{ \omega_a \lambda^2_{a} }{2} \frac{ 2 \beta }{ 1 + \beta })^{ 2 } $.
   It leads to  $ D, 2D, 3D.....$  at $ m=0,1,2.....$ at the $ U(1) $ limit $ \beta=0 $  shown in Fig.\ref{levelevolution}a.
   It is easy to see that at a given $ m-th $ doublet the maximum gap at  $ \alpha=0 $ is smaller than
   the edge gap at $ \alpha=\pm 1/2 $, $ \Delta_{m, U(1)}(\alpha=0 ) <   \Delta_{m, U(1)}(\alpha=\pm 1/2 ) $.
   But both are of the same order.

   Nonetheless, the important phenomena of Goldstone and Higgs mode at the $ U(1) $ limit can still be observed in this $ U(1) $ regime after
   considering these degenerate and non-degenerate perturbations.
   Various photon correlation functions in this $ U(1) $ regime can be evaluated straightforwardly \cite{gold}.

    Now we follow the formation of the bound states just after the $ U(1) $ regime.
    When   $  \frac{ \omega_{-0} }{ D} = \frac{ \pi }{4} \hbar $, namely $ g=g_{c1} $, it just holds the first bound state with
    $ \theta_b = -\theta_a= \frac{\pi}{2} $, $ E_1= 2 \omega_{a} \lambda^{2}_a \frac{ 2 \beta }{ 1+ \beta }$.
    When   $  \frac{\pi }{4} \hbar  < \frac{ \omega_{-0} }{ D} <  \frac{ 3 \pi }{4} \hbar $, it holds the
    first bound state ( Fig.\ref{bound}a ) with energy $ E_1=  2 \omega_{a} \lambda^{2}_a \frac{ 2\beta }{ 1+ \beta } \sin^{2} \theta_{a} $ where
    $ \frac{ \omega_{-0} }{ D} F(\theta_a ) = \frac{\pi }{4} \hbar,
    F(\theta_a ) = \int^{\theta_a}_{0} d \theta \sqrt{ \sin^{2} \theta_a - \sin^{2} \theta },  0 < \theta_a < \pi/2 $.
    It is easy to see $ 0 < F(\theta_a ) < 1, F(\theta_a=\pi/2 ) =1 $.

    When  $  \frac{ \omega_{-0} }{ D } = \frac{ 3 \pi }{4} \hbar $,  namely $ g=g_{c2} $, it just holds the second bound state with
    $ \theta_b = -\theta_a= \frac{\pi}{2} $, $ E_2= 2 \omega_{a} \lambda^{2}_a \frac{ 2\beta }{ 1+ \beta } $.
    While the first bound state energy is given by  $ F(\theta_a )=1/3,  E_1=  2 \omega_{a} \lambda^{2}_a \frac{ 2\beta }{ 1+ \beta } \sin^{2} \theta_a $.
    When  $  \frac{3 \pi }{4} \hbar  < \frac{ \omega_{-0} }{ D} <  \frac{ 5 \pi }{4} \hbar $,
    it holds two bound states ( Fig.\ref{bound}b ) with the energies:
\begin{eqnarray}
   E_1  &= & 2 \omega_{a} \lambda^{2}_a \frac{ 2\beta }{ 1+ \beta } \sin^{2} \theta_{1a},~~ \frac{ \omega_{-0} }{ D} F(\theta_{1a} ) =  \frac{\pi}{4} \hbar
\nonumber  \\
   E_2  & = &  2 \omega_{a} \lambda^{2}_a \frac{ 2\beta }{ 1+ \beta } \sin^{2} \theta_{2a},~~\frac{ \omega_{-0} }{ D} F(\theta_{2a} )  =  \frac{3 \pi }{4} \hbar
\end{eqnarray}
    where one can identify the first atomic energy in Fig.\ref{bound}b:
\begin{equation}
     \epsilon_a= E_2-E_1=2 \omega_{a} \lambda^{2}_a \frac{ 2\beta }{ 1+ \beta } ( \sin^{2} \theta_{2a}-\sin^{2} \theta_{1a} )
\label{epsilona}
\end{equation}
    As expected $ \epsilon_a $ is different than the $ \omega_{-0} $ in Eqn.\ref{atomicm}.

    As $ g \rightarrow \infty $,
    $ D  \rightarrow 0 $,  while $ \omega_{-0} \rightarrow \sqrt{2} \omega_a \sqrt{\frac{ 2\beta }{ 1+ \beta }}  $,
    so the left hand side of Eqn.\ref{boundn} diverges, there are infinite number of bound states shown in Fig.\ref{levelevolution}e.

    Note that because the bound state is either localized around $ \theta_{+} =0 $ or
    $ \theta_{+} = \pi $, so the Berry phase $ \alpha $ in Eqn.\ref{phaselowh} plays no roles,
    so can be dropped. However, as to be shown in the following section, it does play very
    important roles in the quantum tunneling process between the two bound states shown in Fig.\ref{bound} and \ref{instanton}.

\section{ Quantum tunneling between the two bound states: Berry phase and instantons. }
   The instanton solution for a Sine-Gordon model was well known \cite{books}.
   From Eqn.\ref{phaselowh}, we can find the classical instanton solution connecting the
   two minima from  $ \theta_{+} =0 $ or $ \theta_{+} =\pi $: $
   \theta_{+}( \tau ) = 2 tan^{-1} e^{ \omega_{-0} ( \tau -\tau_0) } $
   where $ \tau_0 $ is the center of the instanton.
   Its asymptotic form as $ \tau \rightarrow \infty $ is $
    \theta_{+}( \tau \rightarrow \infty ) \rightarrow \pi -2 e^{-\omega_{-0} ( \tau-\tau_0) } $.
    The corresponding classical instanton  action is:
\begin{equation}
   S_{0}= \frac{ 2 \omega_{-0} }{ D}
\label{scl}
\end{equation}

\begin{figure}
\includegraphics[width=2.8cm]{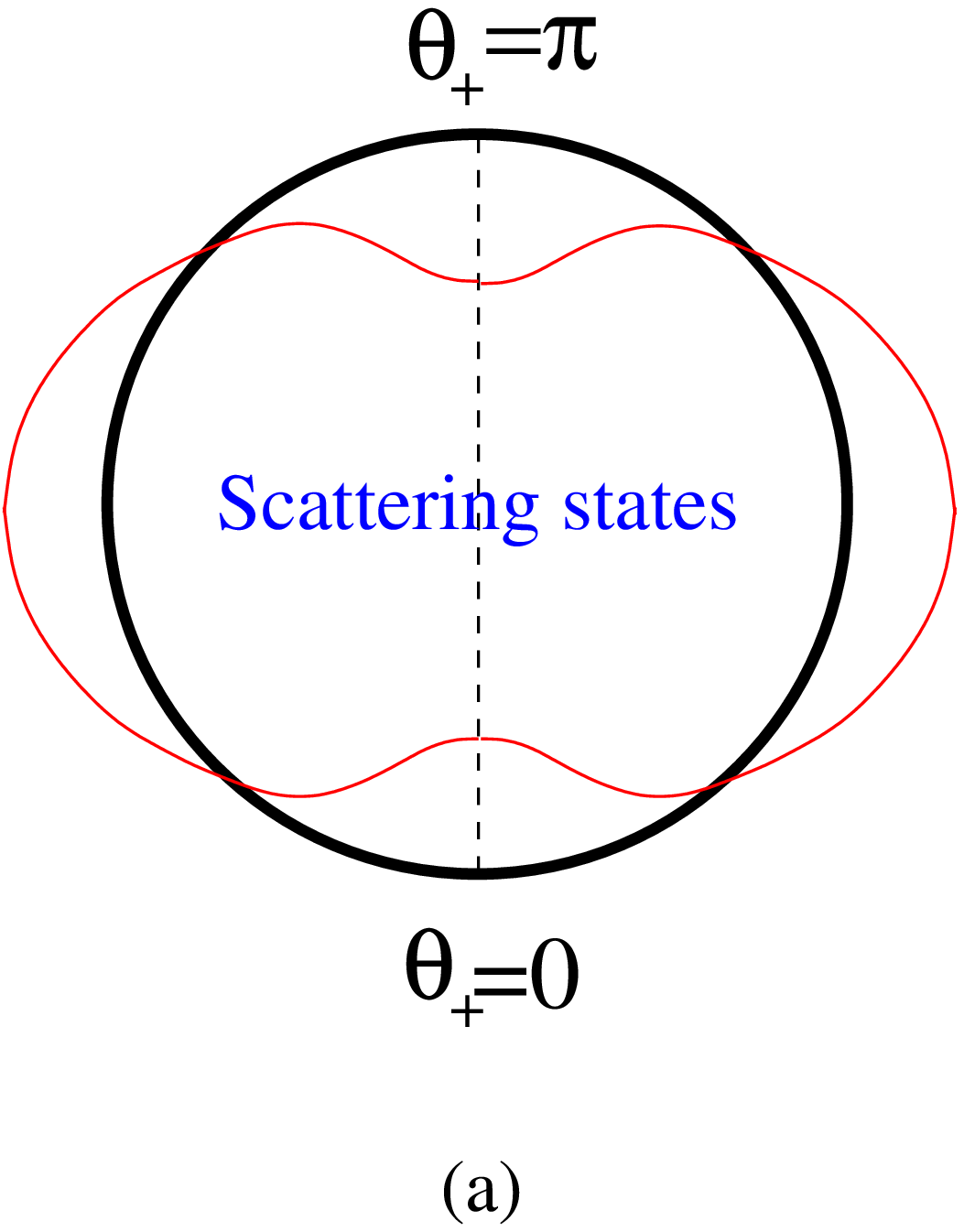}
\hspace{0.1cm}
\includegraphics[width=4.8cm]{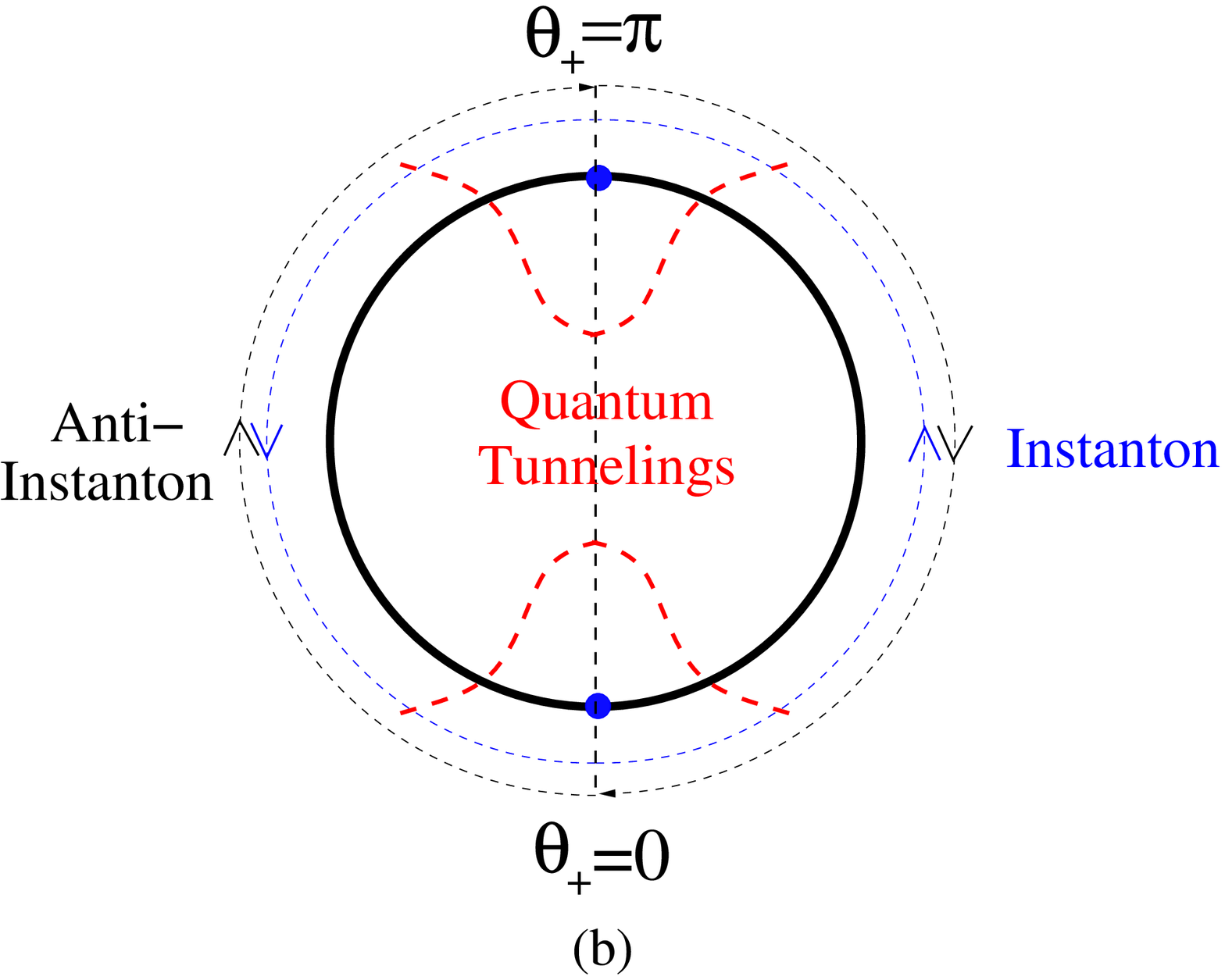}
\caption{ (a) Scattering states: There is a global phase  wandering around the $ \theta_{+} $ circle from $ 0 $ to $ 2 \pi $
with the phase diffusion constant $ D $.
See Fig.\ref{levelevolution}b,Fig.\ref{zeros159}a,b.
(b) Bound states ( Schrodinger cats ): The quantum tunneling process due to the instantons between the two bound states
          at $ \theta_{+}=0 $ and $ \theta_{+} = \pi $.   The counterclockwise ( blue dashed line ) tunneling is an instanton.
          The clockwise ( black dashed line ) tunneling is an anti-instanton. It is the Berry phase which leads to the oscillations of parities in the ground state and excited states. See Fig.\ref{levelevolution}c,d,e,Fig.\ref{zeros159}c.  }
\label{instanton}
\end{figure}

  The instanton problems in the three well known systems (1) a double well  potential (DWP) in a $ \phi^{4} $ theory (2)
  periodic potential problem (PPP) (3) a particle on a circle (POC) are well documented in \cite{books}.
  The tunneling problem in the present problem is related, but different than all the three systems in the following important ways:.
    (1) the potential  $ V(\theta)= 2 \omega_{a} \lambda^{2}_a \frac{ 2\beta }{ 1+ \beta } ( 1-\cos 2 \theta_{+}) $ in Eqn.\ref{phaselowh}
    is a periodic potential in $ \theta_{+} $. In this regard, it is different than the $ \phi^4 $ theory, but similar to PPP.
    (2) The $ \theta_{+} $ is a compact angle confined in $ 0 < \theta_{+} < 2 \pi $.
     In this regard, it is different than the PPP, but similar to POC.
    (3) There are two minima inside the range $ 0 < \theta_{+} < 2 \pi $ instead of just one.
    In this regard, it is different than the POC, but similar to the $ \phi^4 $ theory.
    So the present quantum tunneling problem is a new  class one.
    Furthermore, it is also very important to consider the effects of Berry phase which change the action of
    instanton to $ S_{int}= S_0 + i \alpha \pi $, that of anti-instanton to $ \bar{S}_{int}= S_0 - i \alpha \pi $ ( Fig.\ref{instanton} )
    where $ -1/2 < \alpha < 1/2 $ is the Berry phase in Eqn.\ref{phaselowh}.

    Taking into account the main differences of the present QT problem from the DWP, PPP and POC studied perviously
    \cite{books}, especially the crucial effects of the Berry phase, we can evaluate the transition amplitude from
    $ \theta_{+}= 0 $ to $ \theta_{+}= \pi $ in Fig.\ref{instanton}:
\begin{widetext}
\begin{eqnarray}
    \langle \pi | e^{-H \tau} | 0 \rangle & = & ( \frac{ \omega_{-0} }{ \pi D \hbar} )^{1/2} e^{ \omega_{-0} \tau/2 }
    \sum_{n_1,n_2} \frac{ ( J K e^{-S_{int} } \tau )^{n_1} }{ n_1 ! } \frac{ ( J K e^{-\bar{S}_{int} } \tau )^{n_2} }{ n_2 ! }
    \delta_{n_1-n_2,odd}                    \nonumber  \\
   & = &
    ( \frac{ \omega_{-0} }{ \pi D \hbar} )^{1/2} e^{ \omega_{-0} \tau/2 } \frac{1}{2}[ e^{ 2 J K \tau e^{-S_0} \cos \alpha \pi }
    - e^{ - 2 J K \tau e^{-S_0} \cos \alpha \pi } ]
\label{tuninst}
\end{eqnarray}
\end{widetext}
    where $ n_1 $ ( $ n_2 $ ) is sum over the instanton  ( anti-instanton ), $ J = ( S_0/2 \pi \hbar)^{1/2} $ is given by the
    instanton action in Eqn.\ref{scl} and $ K $ is the ratio
    of two relevant determinants to remove the zero mode of the instantons due to its center  $ \tau_0 $ listed above Eqn.\ref{scl}.
    It can be shown that $ K= C_0 \omega_{-0} $ where $ C_0=2 $ is extracted from the asymptotic form of
    the instanton solution in the $ \tau \rightarrow \infty $ limit listed above Eqn.\ref{scl}.
    Similarly, in finding the transition amplitude from $ 0 $ back to $ 0 $, one only need to change $ \delta_{n_1-n_2,odd} $ to $ \delta_{n_1-n_2, even} $
    in the first line, consequently, the $ - $ sign to the $ + $ sign in the second line in Eqn.\ref{tuninst}.

    The two transition amplitudes lead to the Schrodinger " Cat " state with even/odd  parity:
\begin{eqnarray}
   |e \rangle_{0,SC}  & =  & \frac{A}{\sqrt{2}}  ( |\theta_{+}=0 \rangle + |\theta_{+}=\pi \rangle ),       \nonumber  \\
   |o \rangle_{0,SC}  & =  & \frac{A}{\sqrt{2}} ( |\theta_{+}=0 \rangle - |\theta_{+}=\pi \rangle ),
\label{evenodd0}
\end{eqnarray}
    where  the overlapping coefficient is $ A^{2} = | \langle x=0 | n=0 \rangle |^2= ( \frac{ \omega_{-0}  }{ \pi \hbar D } )^{1/2} $.
    They have the energy $ E_{e/o}= \frac{ \hbar \omega_{-0} }{2} \mp \Delta_{0}/2 $ with
    the  splitting between them  given by:
\begin{eqnarray}
    \Delta_{0}(\alpha) & = &  8 \omega_{-0} ( \cos \alpha \pi) ( \frac{ \omega_{-0} }{ \pi D})^{1/2} e^{-\frac{ 2 \omega_{-0} }{ D} }
     \nonumber   \\
    & \sim & ( \cos \alpha \pi) \sqrt{N} e^{-cN}
\label{splitting}
\end{eqnarray}
    where one can see that it is  the Berry phase which leads to the oscillation of the gap and parity in Eqn.\ref{evenodd0}.
    It vanishes at the two end points $ \alpha = \pm 1/2 $  and reaches maximin at the middle $ \alpha =0 $.
    Note that the Berry phase $ \alpha $ is defined \cite{berryphase,gold} at a given sector $ P $.
    So Eqn.\ref{evenodd0} has a background parity $ \Pi=(-1)^P $.
    So there is an infinite number of oscillating parities in Eqn.\ref{evenodd0} as $ g $ increases.

    Because the Berry phase effects remain in the given $ P $ sector, extending the results in Ref.\cite{high}, we find
    the splitting in the $ n $-th excited bound states  ( $ n=0,1 $ in Fig.\ref{bound}b ):
\begin{equation}
    \Delta_{n}(\alpha) = \frac{1}{ n ! } ( \frac{ 8 \omega_{-0} }{D})^{n} \Delta_{0}
\label{splittingn}
\end{equation}
    where $ n=0,1,2, \cdots $ with the corresponding $n $-the Schrodinger "Cat " state with even/odd parity and
    the energy $ E_{e/o,n}=  ( n + \frac{1}{2} ) \hbar \omega_{-0} \mp \Delta_{n}/2 $:
\begin{eqnarray}
   |e \rangle_{n,SC}  & =  &  \frac{1}{\sqrt{2}}( |n \rangle_L + | n \rangle_R ),      \nonumber  \\
   |o \rangle_{n,SC}  & =  &  \frac{1}{\sqrt{2}}( |n \rangle_L - | n \rangle_R ),
\label{evenoddn}
\end{eqnarray}
    where $  |n \rangle_{L/R} $ is the $ n-$th bound state in the left(right) well in Fig.\ref{bound}a,b.
    Putting $ n=0 $ and projecting it to the coordinate space at $ x=0 $ recovers Eqn.\ref{evenodd0}
    ( Note that the projection to $ x=0 $ does not work for $ n $ is odd ).
    One can see the higher the bound state in the Fig.\ref{bound}, the larger the splitting is.
    The main difference than the energy level pattern in the $ U(1) $ regime is that all the bound states
    have the $ (e,o),(e,o),\cdots $ ( or $ (o,e),(o,e),\cdots $ ) pattern shown in Fig.\ref{levelevolution}d,e and Fig.\ref{photonsplitting}b.
    This important observation is completely consistent with the results achieved from the strong coupling expansion in \cite{strongED}
    after identifying $ n \sim l $.
    For example, as explained in \cite{strongED}, there is an extra oscillating sign $ (-1)^l $ in Eqn.5 in \cite{strongED}
    achieved from the strong coupling expansion, which is crucial to
    reconcile the results achieved from the two independent approaches !

\section{ Photon, squeezing and number correlation functions }

\begin{figure}
\includegraphics[width=3.2cm]{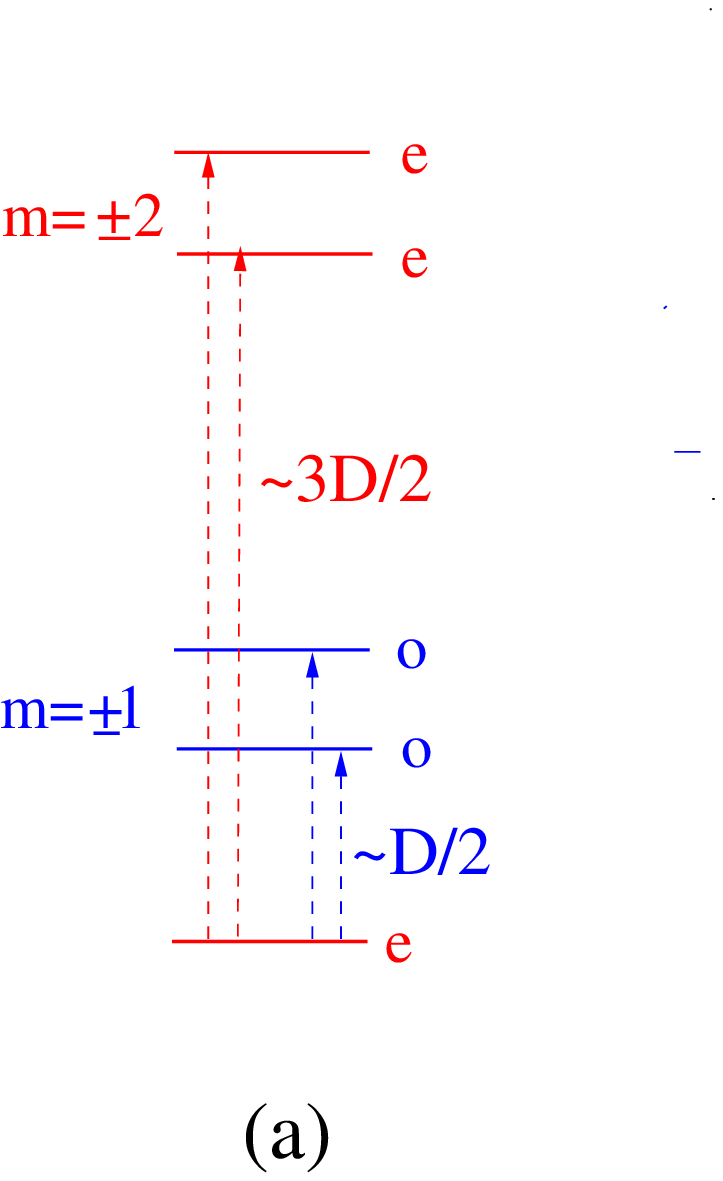}
\includegraphics[width=5cm]{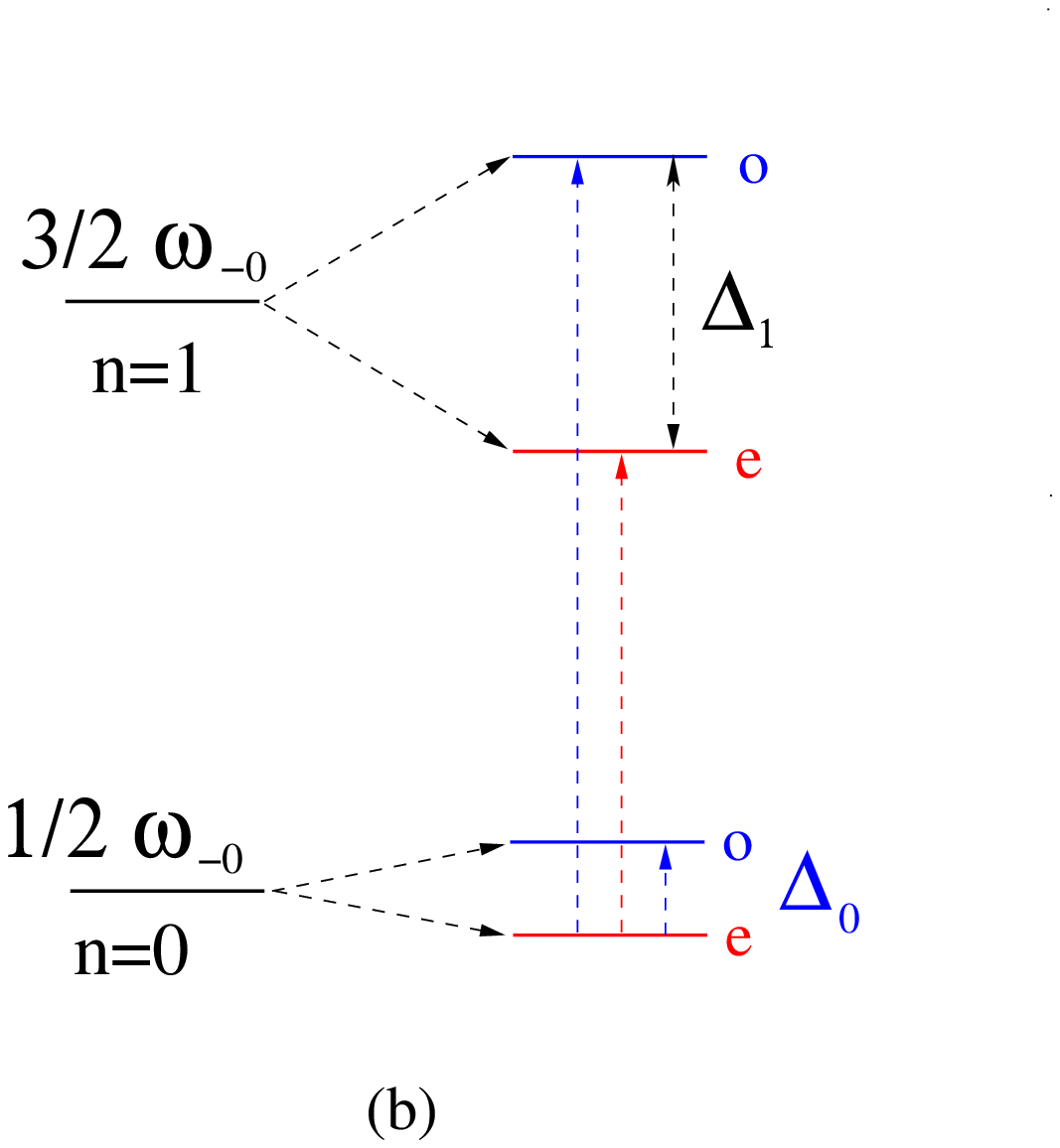}
\caption{ Splittings near the Berry phase $ \alpha=0 $ in the (a) $ U(1) $ regime in Fig.\ref{levelevolution}b and Fig.\ref{instanton}a.
$ D $ is the diffusion constant \cite{berryphase,gold}. $ m=\pm1,\pm2,... $ are the magnetic quantum numbers.
(b) the QT regime in Fig.\ref{levelevolution}d and Fig.\ref{instanton}b.
The ground state $ n=0 $ and the first excited state $ n=1 $ and their splittings $ \Delta_0 $ and
$ \Delta_1 $ due to the quantum tunnelings of instantons subject to the Berry phase.
The blue and red transition lines can be mapped out by photon and photon number correlation functions  respectively.
Shown is the background parity $ (-1)^{P} $ is even (e). When $ (-1)^{P} $ is odd (o), one just changes  even and odd in the figure.}
\label{photonsplitting}
\end{figure}

    Following the procedures  for the $ U(1) $ Dicke model at $ \beta=0 $ in \cite{gold},
    treating the second term in Eqn.\ref{phaselowh} as a small perturbation when $ \beta $ is small,
    using non-degenerate perturbation away from $ \alpha=0 $ and degenerate perturbation near $ \alpha=0 $
    one can evaluate Photon, squeezing and number correlation functions in the $ U(1) $ regime outlined in Fig.\ref{photonsplitting}a.
    Here we focus on calculating these correlation functions in the QT regime outlined in Fig.\ref{photonsplitting}b.

    The above physical pictures in the QT regime inspire us to decompose photon  and atomic operators as:
\begin{equation}
      a= a_L \tau_z,~~~~~ b= - b_L    \tau_z
\label{alising}
\end{equation}
    where $ a_L= \sqrt{ \lambda^{2}_{a} + \delta \rho_a } e^{ i \theta_a},
    b_L=\sqrt{ \lambda^{2}_{b} + \delta \rho_b } e^{ i \theta_b} $ are confined to the left well in Fig.\ref{bound}
    and the $ \tau_z=\pm 1 $ stand for the Left/Right quantum
    wells in Fig.\ref{bound}. It is the $ \tau_z $ component which
    contains the important Berry phase effects and the quantum tunneling process between the Left and Right quantum well
    with the tunneling Hamiltonian $ H_{T}= \Delta_l \tau_x $.
    Note that the two states $ \tau_{z} = \pm 1 $  here is the two Schrodinger Cat states of
     the strongly interacting atom-photon system instead of the
     two levels of an atom $ \sigma_z = \pm 1 $ in the original $ U(1)/Z_2 $ Dicke Hamiltonian Eqn.\ref{u1z2u1}.
    The two energies  $ \Delta_l $ and $ \epsilon_a $  should appear in the single photon correlation function
    ( which contain the magnitude $  \delta \rho_a $,
    the phase $ \theta_a $ and the Ising $  \tau_z $ correlation
    functions as shown in Fig.\ref{crossover}  ) with the corresponding spectral weights $  \sim N, \sim 1 $ respectively.
    By using this decomposition, we will perform the $ 1/J $ expansion to evaluate all the  relevant photon and atomic correlation functions.

    Because the $ Z_2 $ symmetry is broken in either left or right well in Fig.\ref{bound}, one can ignore the periodicity in $ \theta_a, \theta_b $ and expand
    the atom and photon operators as:
\begin{eqnarray}
   a_L & = &  \lambda_a  + i \lambda_a \theta_a +  \frac{ \delta \rho_a }{ 2 \lambda_a } -\frac{ \lambda_a \theta^{2}_{a} }{2} -
   \frac{  (\delta \rho_a)^{2} }{ 8 \lambda^{3}_a } + \cdots     \nonumber   \\
   b_L & = &  \lambda_b  + i \lambda_b \theta_b +  \frac{ \delta \rho_b }{ 2 \lambda_b }  -\frac{ \lambda_b \theta^{2}_{b} }{2} -
   \frac{  (\delta \rho_b)^{2} }{ 8 \lambda^{3}_b } + \cdots
\label{ab}
\end{eqnarray}
   Using the Holstein-Primakoff (HP) representation of the angular momentum operator  $ J_{z}=
   b^{\dagger}b-J, J_{+}= b^{\dagger} \sqrt{ 2 J- b^{\dagger} b}, J_{-}=  \sqrt{ 2 J- b^{\dagger} b} b $, one can evaluate the atomic spin
   correlation functions. Here, we focus on evaluating the photon correlation functions.

    Using Eqn.\ref{phaselowh}, \ref{atomicm}, we can calculate the phase-phase, density-density and
    density-phase correlation functions in the imaginary time $ \tau $:
\begin{eqnarray}
 \langle \theta_{a}(\tau) \theta_{a}(0) \rangle  & =  & \frac{  D }{ 2 \omega_{-0} }  e^{-\omega_{-0} \tau },
 ~ \langle \theta^{2}_{a} \rangle   =   \frac{  D }{ 2 \omega_{-0} }   \nonumber  \\
 \langle \delta \rho_{a}(\tau) \delta \rho_{a}(0) \rangle
 & = & \frac{ \omega_{a} \lambda^{2}_a  }{ 2 \omega_{-0} }\frac{ 2\beta }{ 1+ \beta } e^{-\omega_{-0} \tau },
 ~ \langle  ( \delta \rho_{a} )^{2} \rangle  =  \frac{ \omega_{a} \lambda^{2}_a  }{ 2 \omega_{-0} } \frac{ 2\beta }{ 1+ \beta }   \nonumber  \\
 \langle \delta \rho_{a}(\tau) \theta_{a}(0) \rangle  & = &  -\frac{i}{4} e^{-\omega_{-0} \tau } = - \langle \delta \theta_{a}(\tau) \rho_{a}(0)  \rangle
\label{pluscorrtau}
\end{eqnarray}

   Because the $ \theta_{+} $ is a phase confined on $ 0 < \theta_{+} < 2\pi $, we can define $ q = e^{ i \theta_+} $.
   For $ \theta_{+}=0, \pi $, we can set $ q=  \tau_{z} $ standing for the left/right well in Fig.\ref{bound}
   and find  the correlation function in the $ l-th$ state:
\begin{equation}
    \langle e | \tau_{z}( \tau)  \tau_{z}(0) | e \rangle = e^{ - \Delta_l \tau }
\label{isingtau}
\end{equation}
   where $ \Delta_l $ is the splitting at $ l-th $ state in Fig.\ref{bound} and \ref{photonsplitting}.

     From the decomposition Eqn.\ref{alising}, \ref{ab}, \ref{pluscorrtau} and Eqn.\ref{isingtau}, one can evaluate the photon correlation functions:
\begin{eqnarray}
    \langle a(\tau) a^{\dagger} (0) \rangle   =
    [ \lambda^{2}_a - \frac{ \omega_{-0}}{8 \omega_a}(  \frac{ 1+ \beta }{ 2\beta } )
    - \frac{ \omega_a}{8 \omega_{-0}}( \frac{ 2\beta }{ 1+ \beta } ) ] e^{- \Delta_0 \tau }
                                           \nonumber   \\
    +  \frac{1}{8} [  \frac{ \omega_{-0}}{ \omega_a} ( \frac{ 1+ \beta }{ 2\beta } )
   +   \frac{ \omega_{a}}{ \omega_{-0} }( \frac{ 2\beta }{ 1+ \beta } ) -2 ]   e^{- (\omega_{-0} + (\Delta_1+ \Delta_0 )/2 ) \tau } ~~~~~~
\label{aacorrtun}
\end{eqnarray}
    where the first term containing the ground state splitting $ \Delta_0 $ has the corresponding spectral weight $ \sim N $,
     while the second term containing the  atomic energy $ \epsilon_a $ plus the average of the splittings
     at $ n=0 $ and $ n=1 $ in Fig.\ref{photonsplitting} has the spectral weight $ \sim 1 $.

     Very similarly, one can evaluate the anomalous  photon correlation functions:
\begin{eqnarray}
    \langle a(\tau) a(0) \rangle   =
    [ \lambda^{2}_a - \frac{ \omega_{-0}}{8 \omega_a}(  \frac{ 1+ \beta }{ 2\beta } )
    - \frac{ \omega_a}{8 \omega_{-0}}( \frac{ 2\beta }{ 1+ \beta } ) ] e^{- \Delta_0 \tau }
                                           \nonumber   \\
    +  \frac{1}{8} [    \frac{ \omega_{a}}{ \omega_{-0} }( \frac{ 2\beta }{ 1+ \beta } ) - \frac{ \omega_{-0}}{ \omega_a} ( \frac{ 1+ \beta }{ 2\beta } )   ]   e^{- ( \omega_{-0} + (\Delta_1+ \Delta_0 )/2  ) \tau } ~~~~~~
\label{aacorrtuna}
\end{eqnarray}
     where the first term containing the ground state splitting $ \Delta_0 $ has the same spectral weight $ \sim N $ as its counterpart in Eqn.\ref{aacorrtun},
     while the second term containing the  atomic energy $ \epsilon_a $ plus the average of the splittings
     at $ n=0 $ and $ n=1 $ in Fig.\ref{photonsplitting} has a different spectral weight $ \sim 1 $, maybe even different sign
     than its counterpart in Eqn.\ref{aacorrtun}.

     One can also compute the photon number correlation functions:
\begin{equation}
  \langle n(\tau) n(0) \rangle - \langle n \rangle^2=
  \frac{ \omega_{a} \lambda^{2}_a  }{ 2 \omega_{-0} }\frac{ 2\beta }{ 1+ \beta }  e^{- ( \omega_{-0} - (\Delta_1- \Delta_0 )/2  ) \tau }
\label{nn1}
\end{equation}
    where $ \langle n \rangle =\lambda^{2}_a $ is the photon number at the ground state.
    It contains the atomic energy $ \epsilon_a $ minus the difference of the splittings
    between $ n=1 $ and $ n=0 $ in Fig.\ref{photonsplitting} and has a  spectral weight $ \sim N $.

     So all the parameters of the cavity systems such as the doublet splittings $ \Delta_0( \alpha), \Delta_1( \alpha) $ and
     the atomic energy $ \epsilon_a $ are encoded in the photon normal Eqn.\ref{aacorrtun} and
     anomalous Green function Eqn.\ref{aacorrtuna}  and photon number correlation function
     Eqn.\ref{nn1}. They can be measured by photoluminescence,
     phase sensitive homodyne and  Hanbury-Brown-Twiss ( HBT ) type of experiments \cite{exciton} respectively.



\vspace{0.2cm}
\section{ Comparison with the results from the Exact Diagonization and the strong coupling expansion. }

\begin{widetext}

\begin{figure}
\includegraphics[width=8cm]{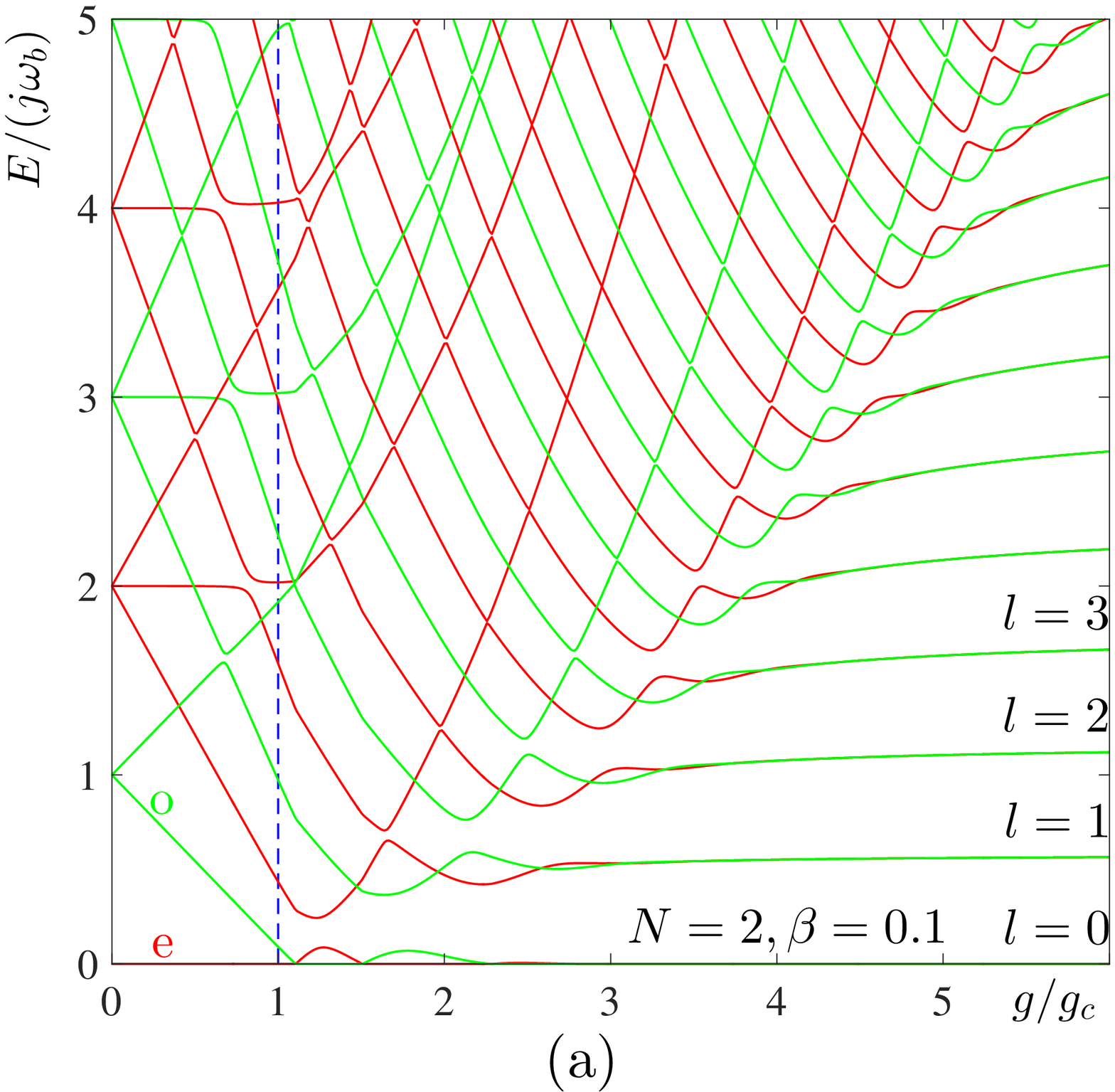}
\hspace{0.1cm}
\includegraphics[width=8cm]{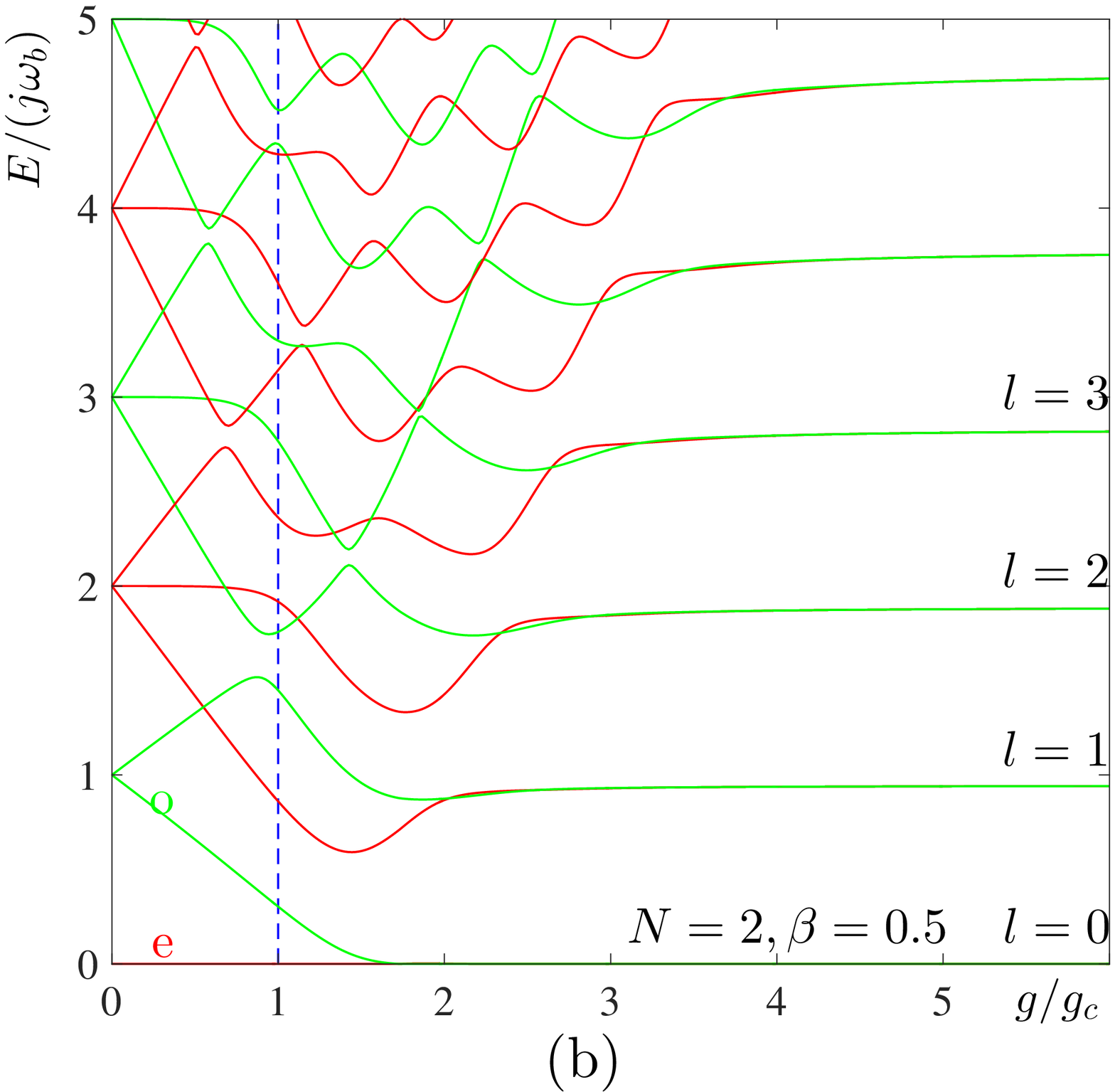}
\includegraphics[width=8cm]{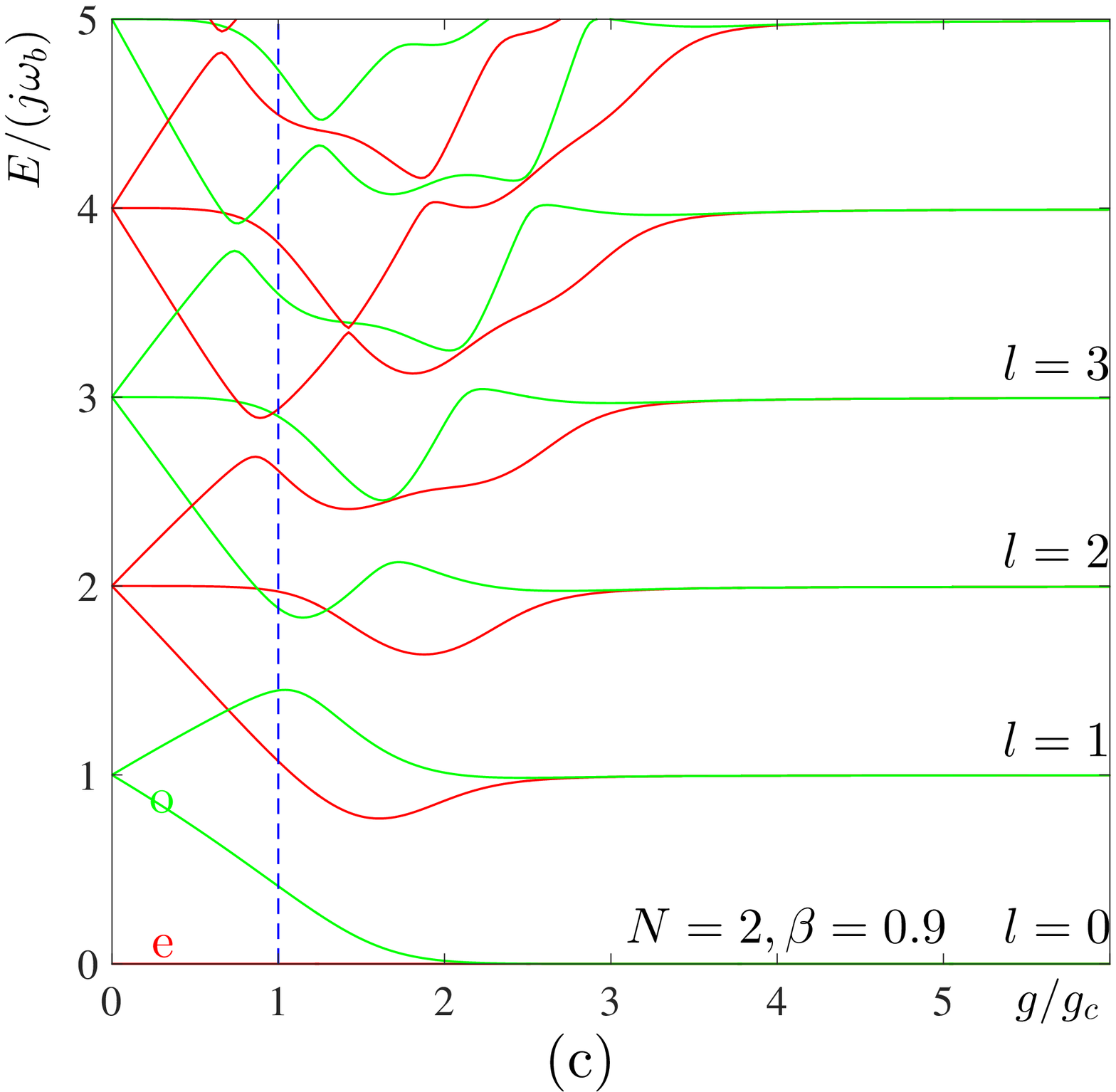}
\hspace{0.1cm}
\includegraphics[width=8cm]{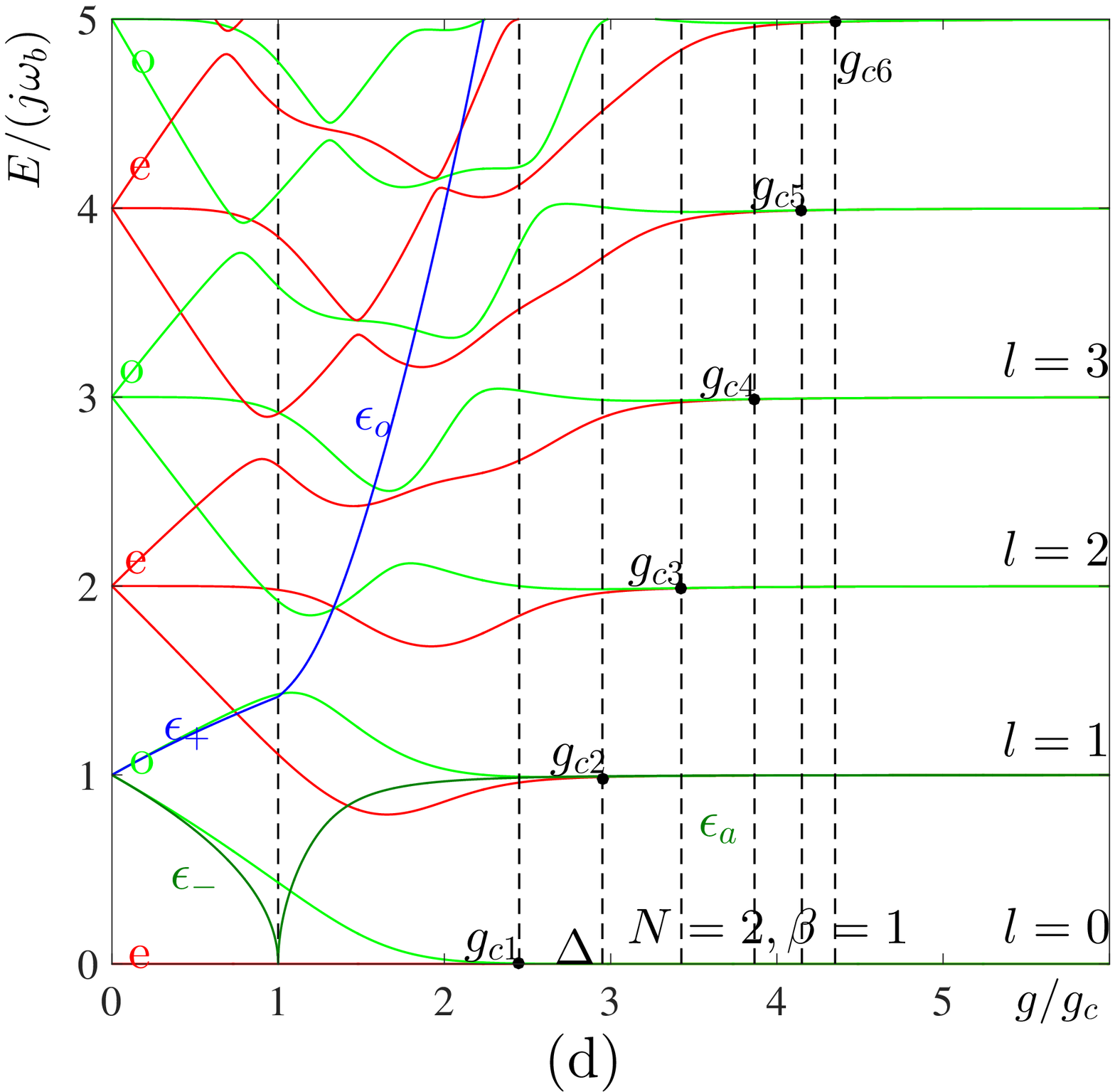}
\caption{ ED results for the energy levels at $ N=2, \beta=0.1, 0.5, 0.9, 1 $ for (a)-(d) respectively.
For simplicity, we only show $ \omega_a=\omega_b $ case.
The parity even (e) and odd (o) are indicated.
There are none, one and two level crossing(s) in the normal regime at $ l=0,1,2 $ respectively.
(a) For $ \beta=0.1 $, there are considerable ranges of the $ U(1) $ regime $ g_c < g < g_{c1} $ before the first bound state formation
at $ l=0 $, then it evolves to the QT regime where the bound states start to form at $ g_{c1}, g_{c2},g_{c3}.....$ ( which are labeled only in (d), but could be labeled in (a)-(c) also )
at $ l=0,1,2....$ as $g/g_c $ increases. There is a one to one correspondence to Fig.\ref{levelevolution}.
(b) For $ \beta=0.5 $, the $ U(1) $ regime becomes much smaller.
(c) For $ \beta=0.9 $, the $ U(1) $ regime disappears.
When expanding the doublets at $ l=0,1,....$, as $ g/g_c $ increases, there are infinite energy level crossings leading to
the oscillations of parities at the ground states at  $ l=0,1,....$ manifolds shown in Fig.\ref{zeros159}.
As $ \beta \rightarrow 1^{-} $, all the zeros are pushed to infinity.
(d) The $ Z_2 $ limit $ \beta=1 $.
There are consecutive energy level mergings at $ g_c < g_{c1} < g_{c1} < \cdots $  which signify the bound state formations at
$ l=0,1,2....$ and the energy levels become approximately {\sl flat }.
There are no $ U(1) $ regime, no level crossing between the even and odd parity pairs.
The ground state pattern is $ (e,o),(e,o),\cdots $ at $ l=0,1,2....$.
Only the atomic energies at $ l=0,1,2....$ are labeled. As $ g/g_c \rightarrow \infty $ limit, they approach to $ l \omega_a $ from below \cite{strongED}.}
\label{leveledn2}
\end{figure}

\end{widetext}

\begin{widetext}

\begin{figure}
\includegraphics[width=5.0cm]{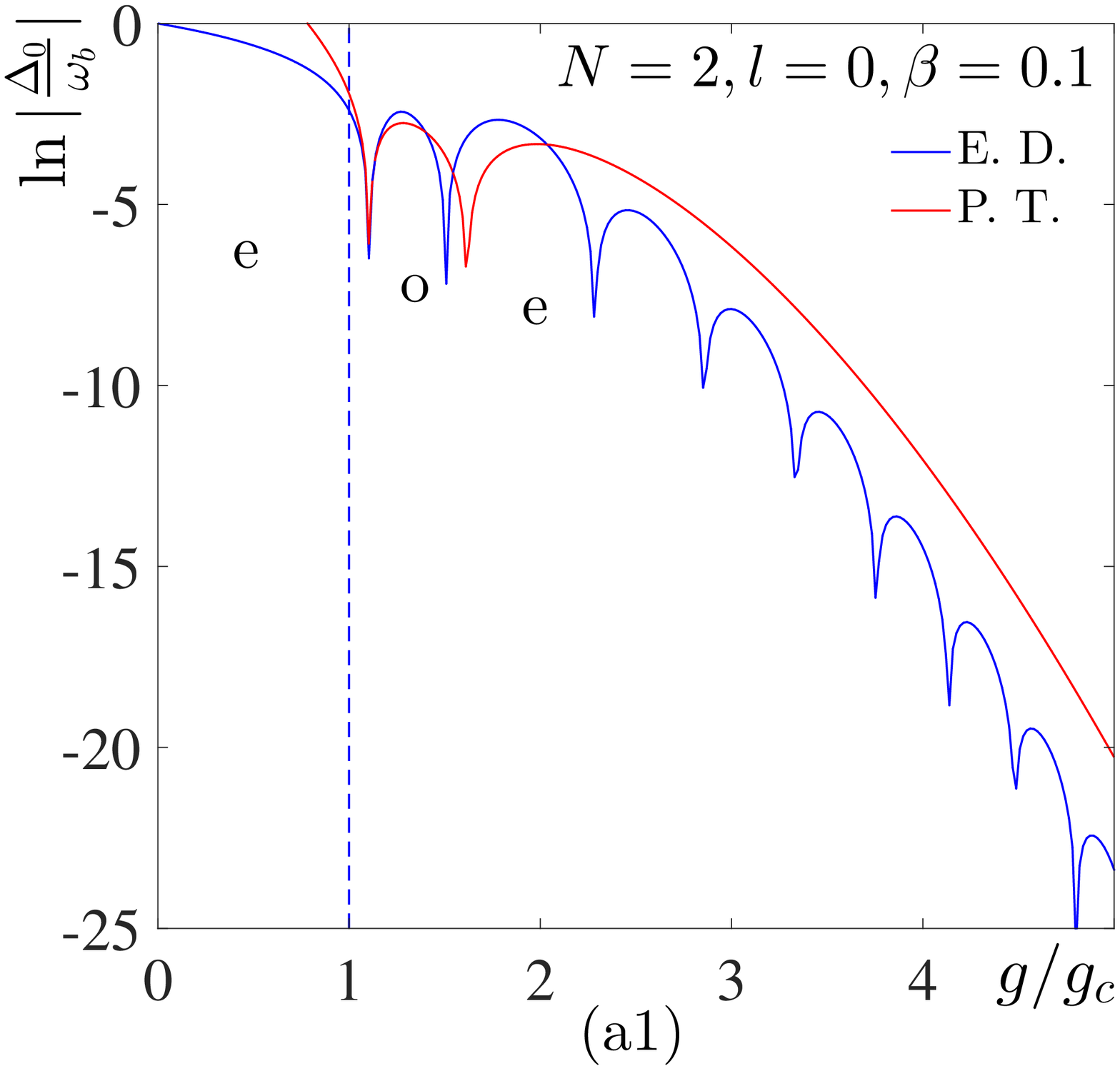}
\hspace{0.2cm}
\includegraphics[width=5.0cm]{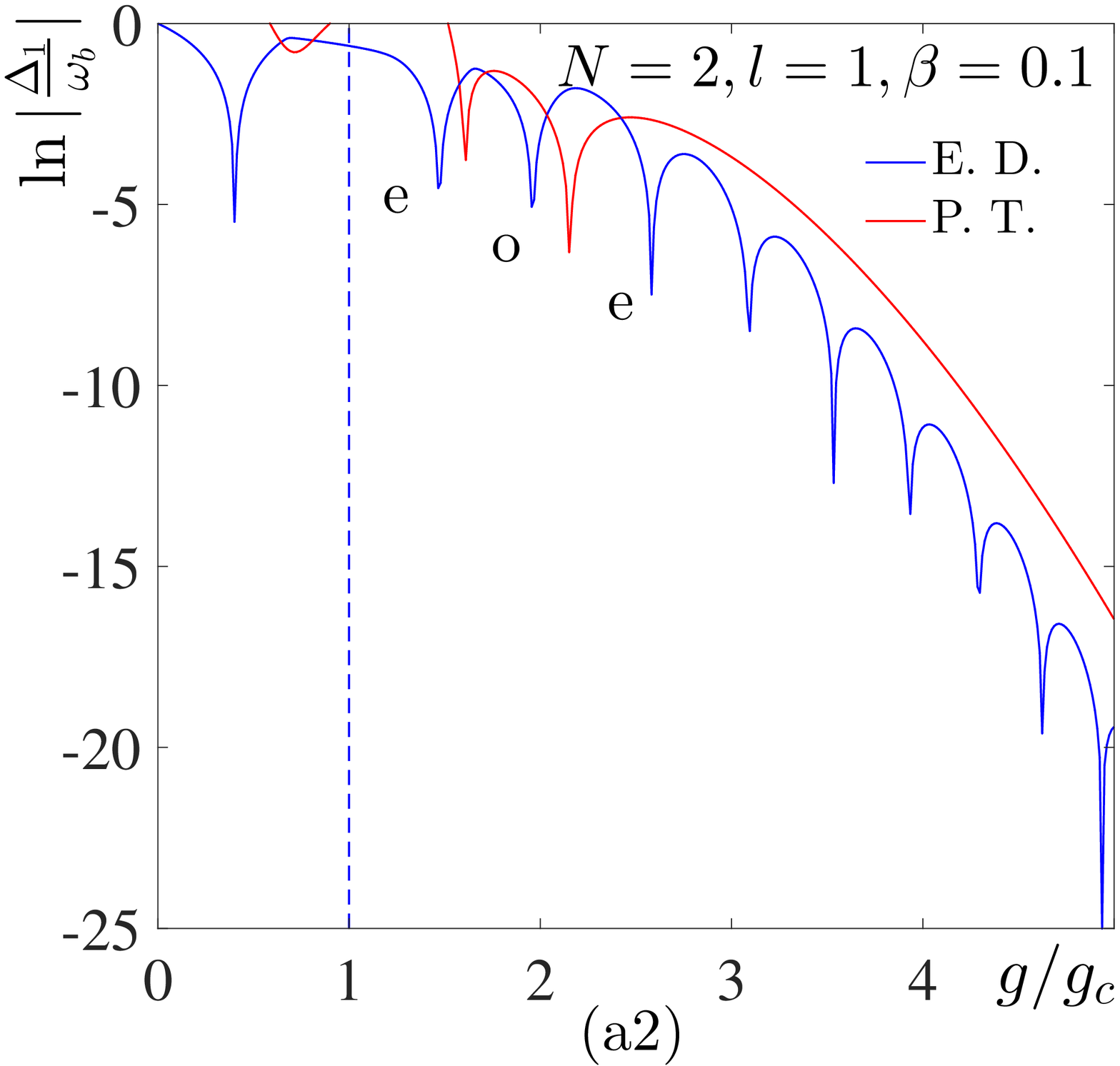}
\hspace{0.2cm}
\includegraphics[width=5.0cm]{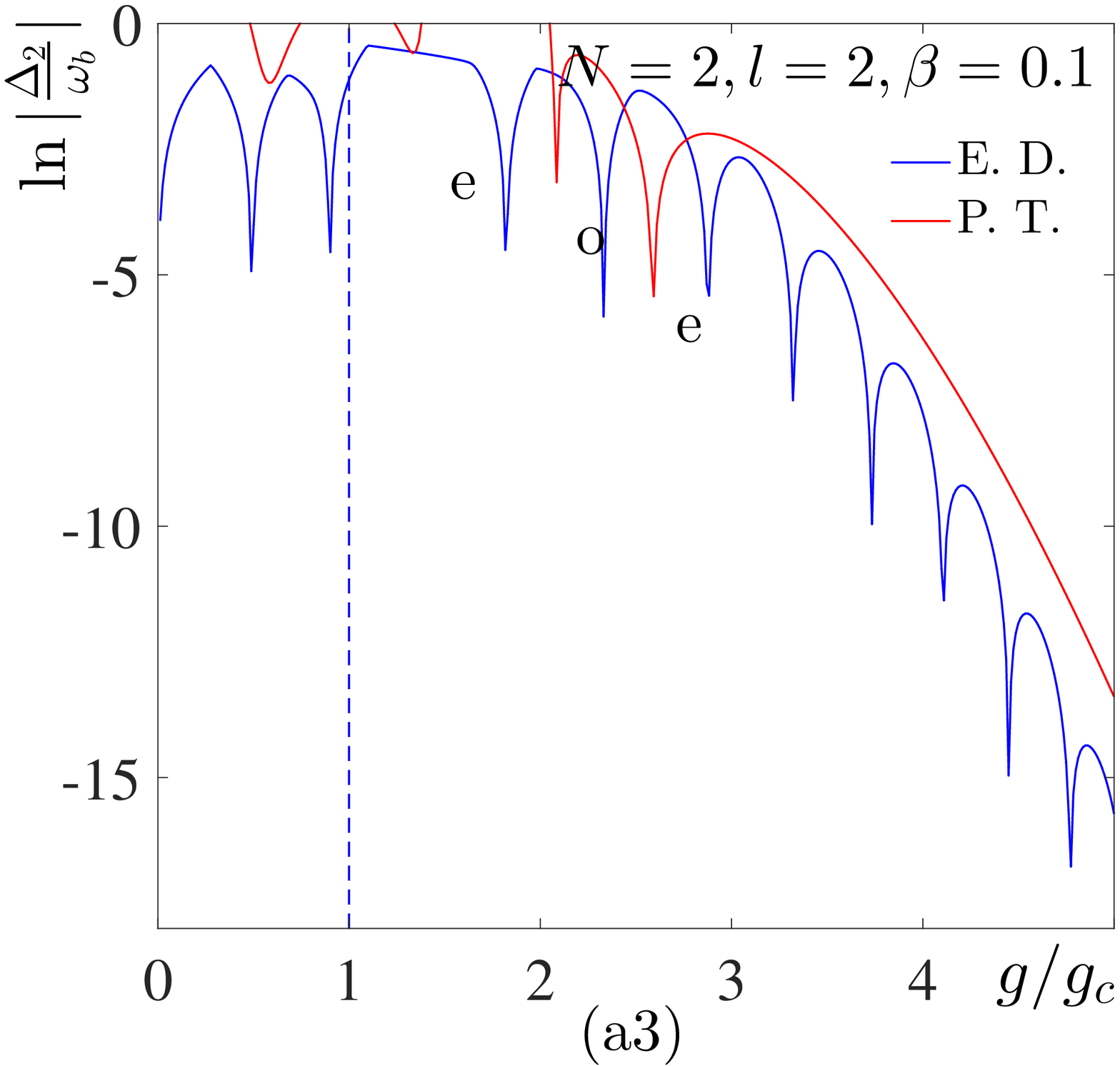}
\includegraphics[width=5cm]{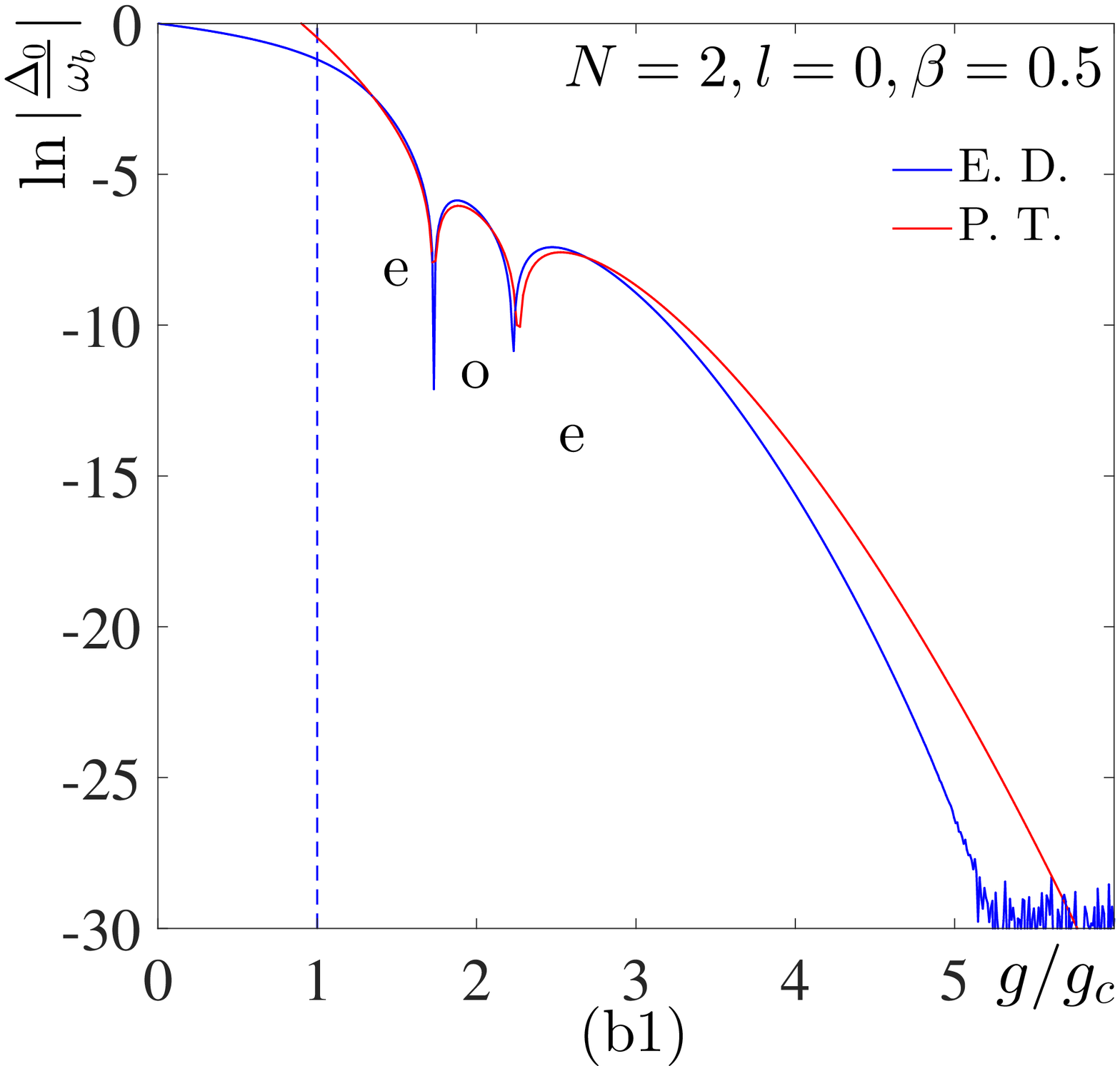}
\hspace{0.2cm}
\includegraphics[width=5cm]{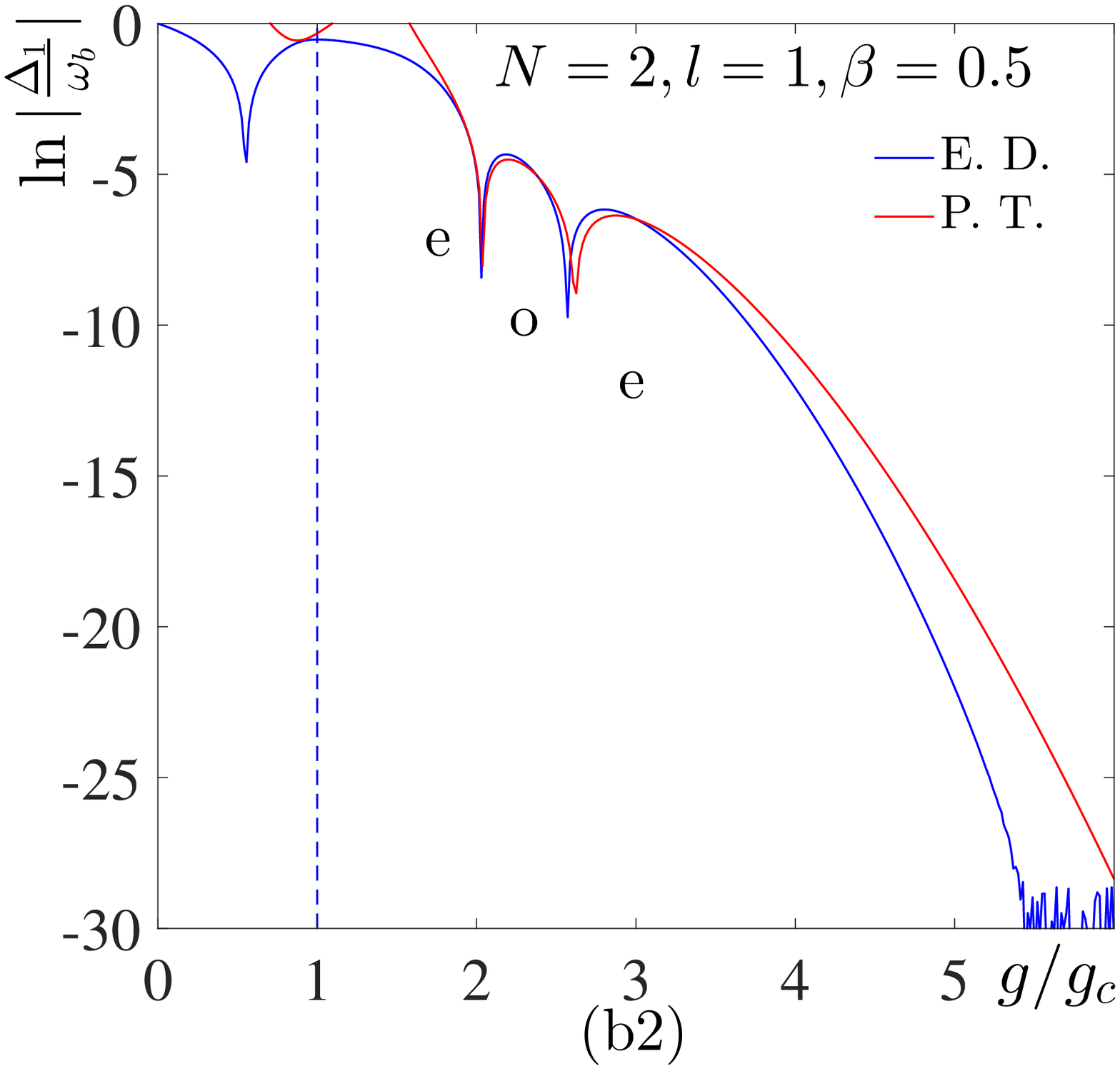}
\hspace{0.2cm}
\includegraphics[width=5cm]{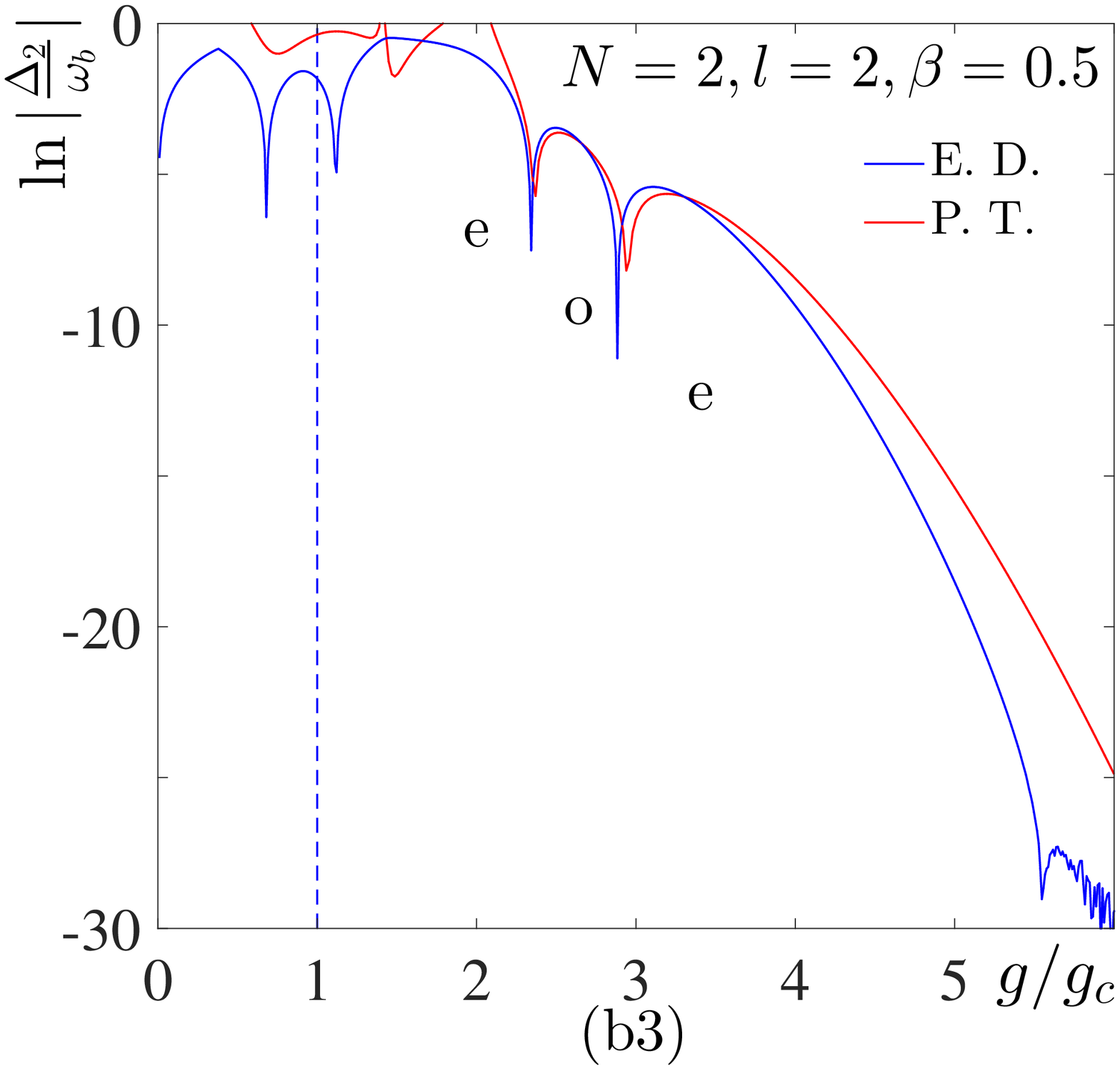}
\includegraphics[width=5cm]{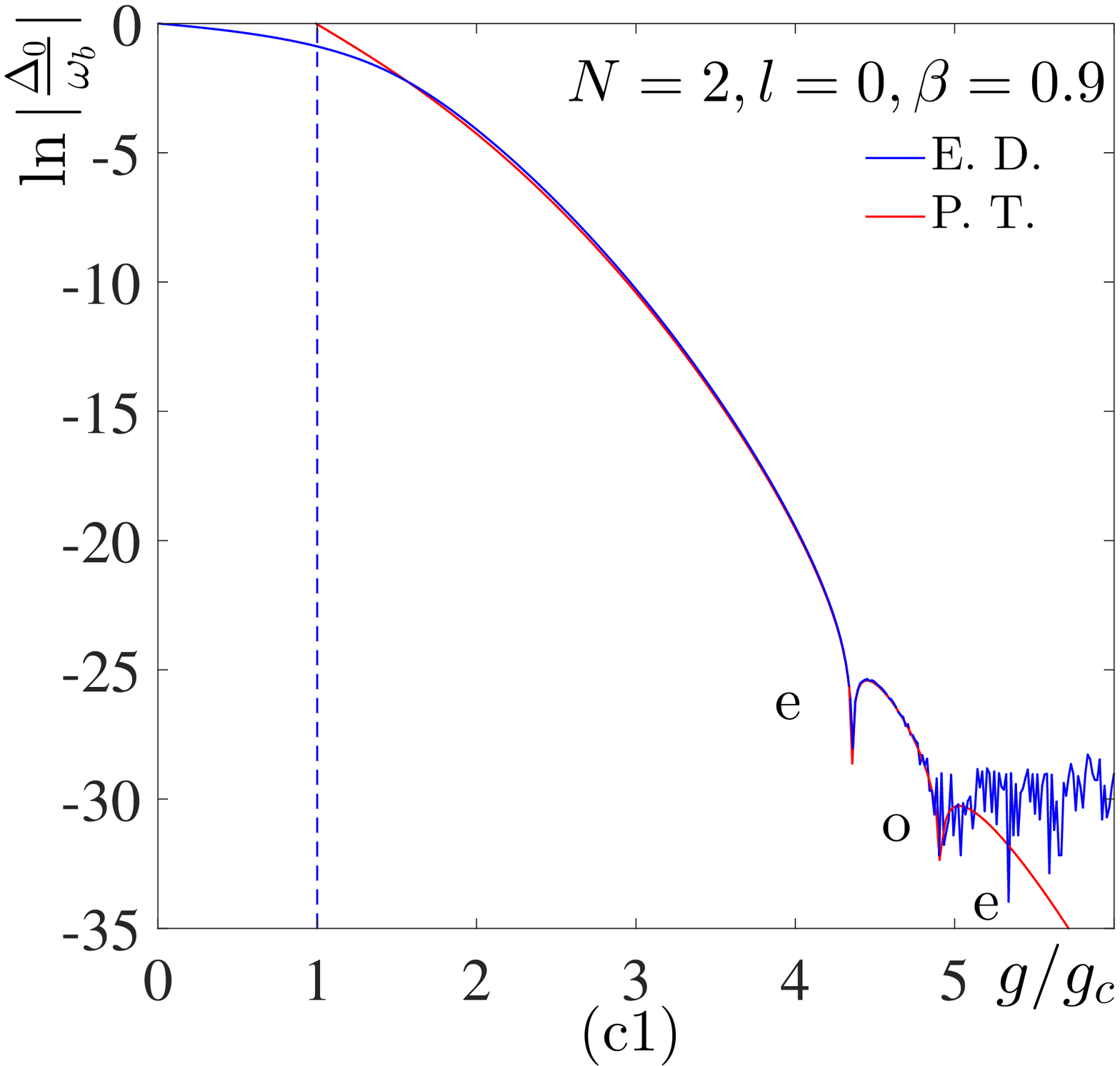}
\hspace{0.2cm}
\includegraphics[width=5cm]{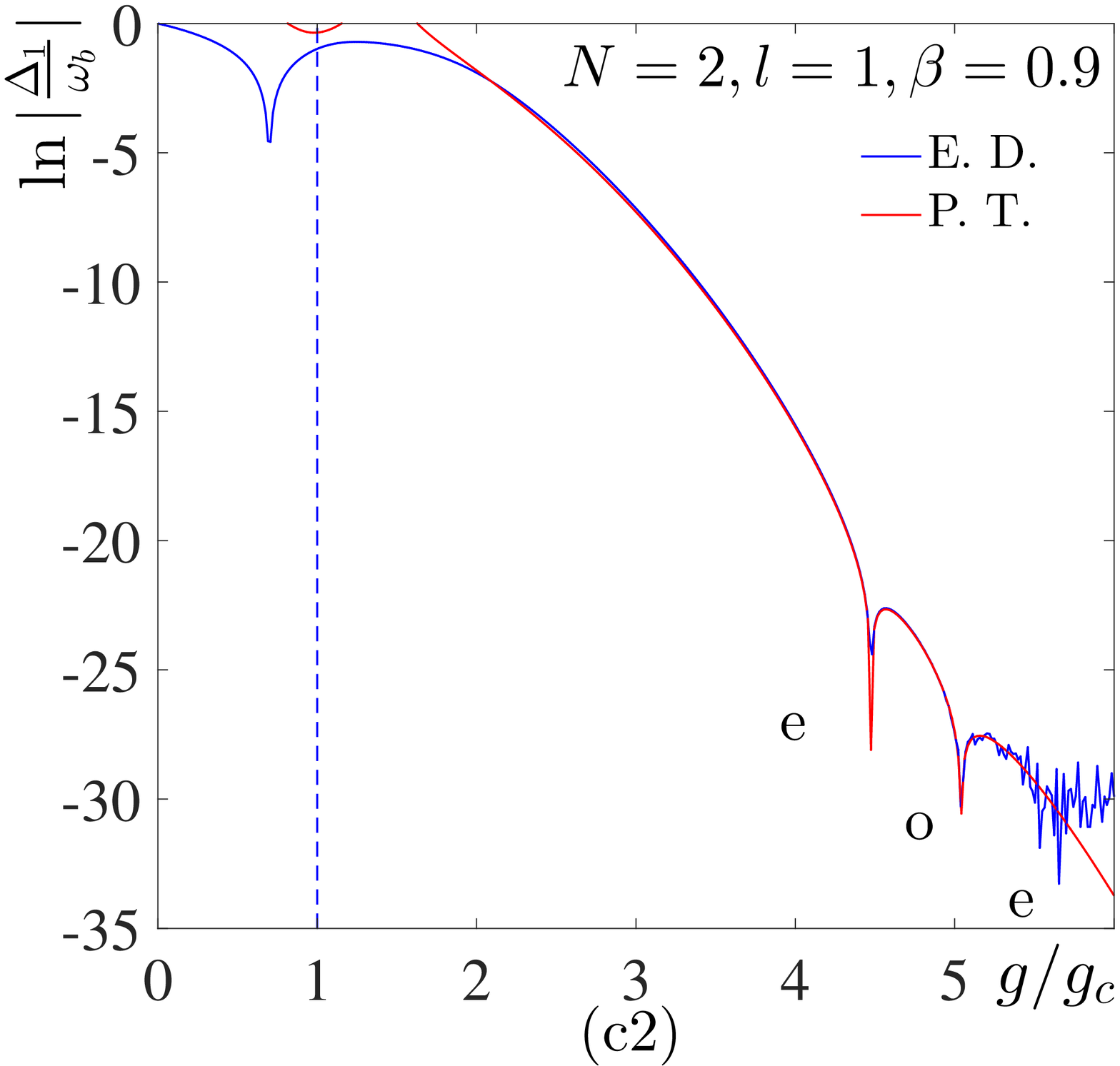}
\hspace{0.2cm}
\includegraphics[width=5cm]{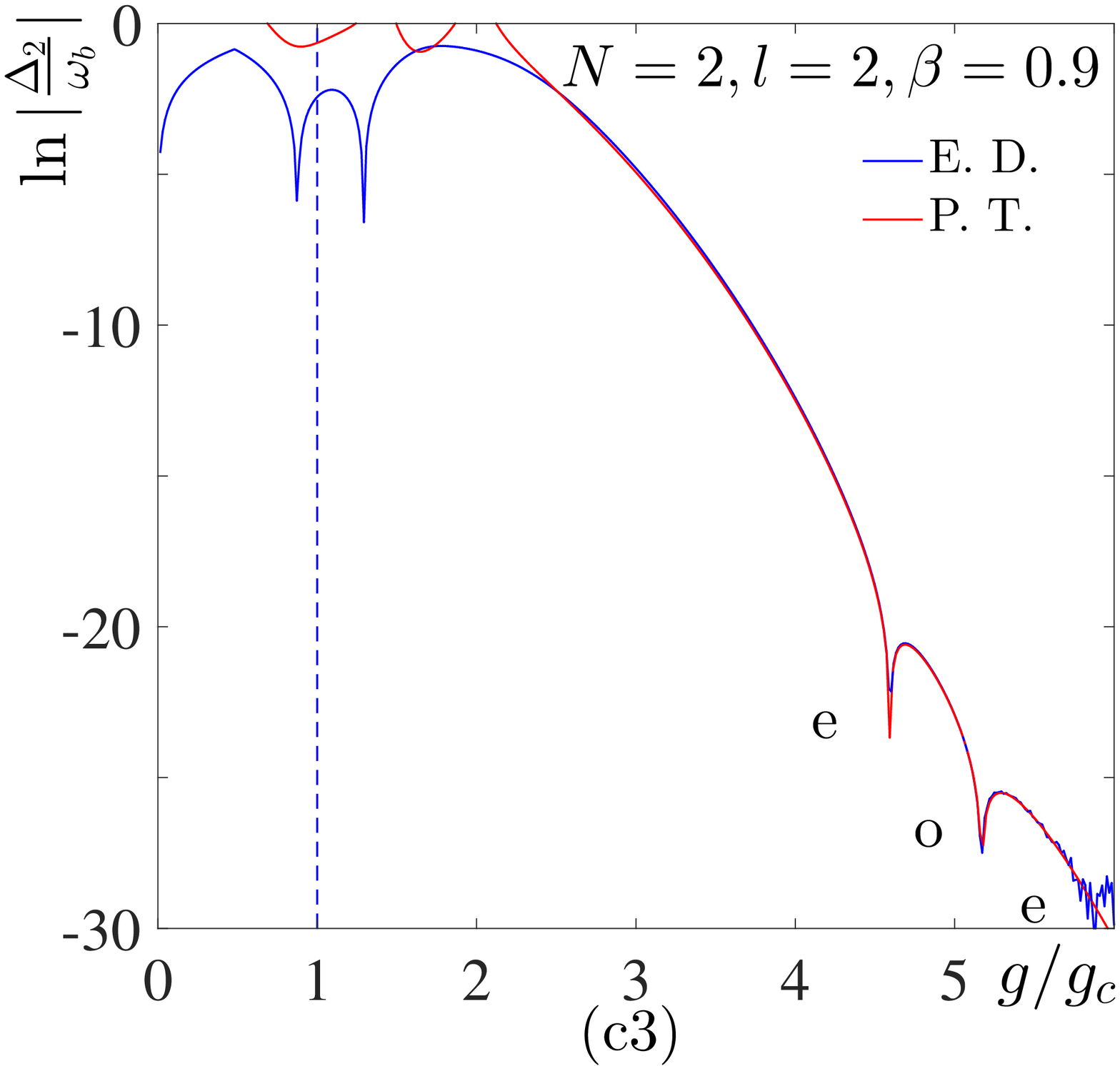}
\caption{  Identify $ U(1) $ and QT regimes in the results achieved from the strong coupling expansion and the ED \cite{strongED}.
In the Log scale, the even/odd splitting $ \Delta_l $ for $ N=2 $ at different $ \beta=0.1, 0.5,0.9 $ in the row and
at $ l=0, 1, 2 $ in the column calculated by strong coupling expansion ( red line ) versus that by the ED ( blue line ) \cite{strongED}.
The labels $ e $ and $ o $ are the parity of the ground states.
 The first even parity is in the normal
 regime in (a) and (b), but becomes a bound state at $ g_{c1} $ which is before the first zero in (c). The next one must be the odd parity state at $ l=0,1,2 $.
 Note also the changes of the numbers in the vertical axis at the  Log scale telling the splitting increases
 as $ l=0 $ to $ l=1 $, then to $ l=2 $ as dictated by Eqn.\ref{deltau1},\ref{deltau1m} for the scattering states in (a) and (b)
 and by Eqn.\ref{splitting} and \ref{splittingn} for the bound states in (c).
 There are also none, one and two level crossing(s) in the normal regime at $ l=0,1,2 $ respectively.
 Eqn.\ref{splitting} dictates there are infinite number of zeros.
 The strong coupling expansion in \cite{strongED} to $ N-$th order only gives the first $ N $ zeros.
 The ED gives infinite number of zeros ( but not shown here ) after the first $ N=2 $ zeros which can only
 be achieved from higher order perturbation calculations in the strong coupling expansion.
 (a) The strong coupling results  \cite{strongED} match well with those from ED at $ l=0 $, but not too well at $ l=1,2 $ in the first
 $ N=2 $ zeros at $ \beta=0.1 $ in the $ U(1) $ regime. Even so, they seem match well the envelop of the splitting
 at $ l=0,1,2 $ ( namely, the maximum splitting at $ \alpha=0 $ ).
 This causes no concerns because the agreement is not expected in the $ U(1) $ regime too close to $ g_c $ when $ \beta $ is small.
 Compare with Fig.\ref{leveledn2}a, the first $ N=2 $ zeros at $ l=0,1,2 $ are due to the scattering states in the $ U(1) $ regime
 in Fig.\ref{levelevolution}b. (2) The strong coupling results  \cite{strongED} match very well with those from ED in the first
 $ N=2 $ zeros even at $ \beta=0.5 $.
 The other zeros are far apart from the first $ N=2 $ zeros and out of the scope in the figure.
 Compare with Fig.\ref{leveledn2}b,  the first $ N=2 $ zeros at $ l=0,1,2 $ are still due to the scattering states
 in the narrow $ U(1) $ regime in Fig.\ref{levelevolution}b. Observe that the strong coupling expansion works
 very well  and reproduces precisely the Berry phase effects in the $ U(1) $ regime not too close to $ g_c $.
 As shown in Fig.\ref{levelevolution}b, there are
 shifts of zeros to the right if one follow the ground state with the odd parity.
(3) The match is essentially perfect in the first $ N=2 $ zeros at $ \beta=0.9 $  in the QT regime.
 The other zeros are far apart from the first $ N=2 $ zeros and out of the scope in the figure. At too strong couplings, the ED may become ( noise ) un-reliable due to the cutoff introduced in the ED. Compare with Fig.\ref{leveledn2}c, all the zeros are well beyond $ g_{c1}, g_{c2}, g_{c3}...$ in Fig.\ref{leveledn2}c,
 so the first $ N=2 $ zeros are the first two bound states
 at $ l=0,1,2 $ in the QT regime in Fig.\ref{crossover} and \ref{levelevolution}e. The $ U(1) $ regime is squeezed out.
 Their locations are nearly independent of $ l $ as dictated by Eqn.\ref{splittingn}  for the bound states
 satisfying $ G = \frac{g}{g_c} \frac{1}{\sqrt{N} } \gg 1 $.  }
\label{zeros159}
\end{figure}

\end{widetext}

 When $ \beta \neq 1 $ in Eqn.\ref{u1z2u1}, it is not convenient to perform the ED in the coherent basis anymore used in \cite{china},
so we did the ED in the orginal ( Fock )  basis. In the Fock space, the complete basis is $ | n \rangle | j, m \rangle, n=0,1,2,.....\infty, j=N/2,
 m=-j,.....,j $  where the $ n $ is the number of photons and the $ | j, m \rangle
 $ is the Dicke states. In performing the ED, following \cite{chaos}, one has to use a
 truncated basis $ n=0,1,......n_c $ in the photon sector  where the $ n_c \sim 100 $ is the maximum photon number in the artificially truncated Hilbert space.
 As long as the low energy levels in Fig.\ref{leveledn2} and Fig.\ref{zeros159} are well below $ n_c \omega_a $, then the
 energy levels should be very close to the exact results without the truncation ( namely, sending $  n_c \rightarrow \infty $ ).
 However, the ED may not be precise anymore when $ g $ gets too close to the upper cutoff introduced in the ED calculation
 as shown in Fig.\ref{zeros159}c.

In Fig.\ref{leveledn2}, we show the ED results for the energy levels for $ N=2 $ at $ \beta=0.1,0.5,0.9,1$.
It matches precisely the theoretically predicted energy level evolutions shown in Fig.\ref{levelevolution}.
At $ N=2 $, when $ \beta < \beta_{c} \sim 0.6 $ ( which, in fact, only weakly depends on $ N $ ),
there is always a $ U(1) $ regime  Fig.\ref{levelevolution}b,c before the formations of bound states in the QT regime in Fig.\ref{leveledn2}a,b.
It is the Berry phase which leads to the parity oscillations in both regimes.
However, $ \beta > \beta_{c} $, the systems get to the formations of bound states directly in Fig.\ref{leveledn2}c.
It is still the Berry phase which leads to the parity oscillations in the QT regime after the formations of bound states as shown in Fig.\ref{instanton}b.
At $ \beta =1 $, the Berry phase effects and the level crossings are pushed to infinity, so no parity oscillations anymore in Fig.\ref{leveledn2}d.

   The important relation Eqn.\ref{splittingn} takes the same form as Eqn.5 in \cite{strongED} except the absence of the extra $ (-1)^{l} $ factor.
   As said at the end of Sec.IV, it is this absence of extra $ (-1)^{l} $ which reconciles the results achieved from the two independent approaches.
   Eqn.\ref{splitting} indicates that there are infinite number of zeros due to the Berry phase interference effects in the instanton tunneling process  in Fig.\ref{instanton}.  Eqn.\ref{splittingn} indicates that the positions of the zeros are independent of $ n $.
   This is indeed confirmed by the ED shown in Fig.\ref{zeros159}c
   for $ N=2, \beta=0.9, l=0,1,2 $ where the positions of the first $ N=2 $ zeros only depend on $ l $ very weakly.
   So between the two zeros, at $ l=0,1,2,\cdots $, the energy levels are either in the pattern
   $ (e,o),(e,o),\cdots $  or $ (o,e),(o,e),\cdots $ in Fig.\ref{levelevolution}d and Fig.\ref{photonsplitting}b.
   This important result completely substantiates
   the results achieved from the strong coupling expansion in \cite{strongED}.
   The fact that the same fantastic phenomena are reached from two independent analytic approaches, then confirmed by ED
   indicates that the results are correct, independent of the $ 1/J $ expansion or strong coupling expansion we made.

\section{ Experimental detection of the Berry phase effects of the instanton tunneling events  }
  There have been extensive efforts to realize the Schrodinger Cat state
  in trapped ions \cite{cat1} and superconducting qubit systems \cite{cat2}.
  Here, the Schrodinger Cats in Eqn.\ref{evenoddn} with $ n=0,1,2...$  can be
  prepared in the QT regime in Fig.\ref{crossover}a, its size can be continuously tuned from $ N \sim 3-9 $,
  it involves all the $ N $  number of atoms ( qubits ) and photons strongly coupled inside the
  cavity and could have
  important applications in quantum information processions.

  With $ N \sim 10^{5} $ atoms of $ ^{87}Rb $ inside a cavity \cite{orbitalt,orbital,switch}, the system is
  essentially in the thermodynamic limit of the $ Z_2 $ Dicke model,
  so the novel physical phenomena in the QT regime in Fig.\ref{crossover} at finite small $ N $
  explored in this work are hard to observe.
  The experiment \cite{switch}
  first adiabatically prepared the system in one of the two bound states in  Fig.\ref{bound}b by applying a small $ Z_2 $ symmetry breaking field,
  then {\bf turn off and on} the transverse pumping laser in Fig.\ref{exp}a and
  observed the coherent switch with the frequency $ \omega_b $ between the two ground states in Fig.\ref{bound}b by an optical heterodyne detection.
   As emphasized in this work, in order to observe the Berry phase interference effects, one has to move away from the $ Z_2 $ limit
   realized in the experiments \cite{orbitalt,orbital,switch}, namely, $ 0 < \beta < 1 $. This has been realized in the recent experiment
   \cite{expggprime} which can tune $ \beta $ from $ 0 $ to $ 1 $.
   With the recent advances of manipulating a few atoms \cite{fewboson,fewfermion}, the number of atoms can be reduced to a few to a few hundreds,
   then the $ U(1) $ and the QT regime in Fig.\ref{crossover}, Fig.\ref{bound}a,b span a large parameter regimes.
   One can first adiabatically prepare the system in the left or right bound states in Fig.\ref{bound} with $ n=0,1,... $,
   but still {\bf keep } the transverse pumping laser in Fig.\ref{exp}a,
   then observe by the optical heterodyne detection \cite{switch} the coherent oscillation probability between the two bound states:
\begin{equation}
    P(\alpha, t)=\cos \Delta_n(\alpha) t
\end{equation}
     where the $ \Delta_n(\alpha) $ is given by Eqn.\ref{splittingn}.



   In circuit QED systems, there are various experimental set-ups such as charge,
   flux, phase qubits or qutrits, the couplings could be capacitive or inductive through $ \Lambda, V, \Xi $ or the $ \Delta $ shape \cite{you}.
   Especially, continuously changing $ 0 < \beta < 1 $ has been achieved in the recent experiment \cite{qubitstrong}.
   An shown in \cite{gold}, by tuning the potential
   scattering term $  \lambda_z  J_{z} a^{\dagger} a/j $ and the qubit-qubit
   interaction term $ u J^{2}_{z}/j $, the critical coupling  $g_c$  in Fig.\ref{crossover} is reduced to $  g_c = \sqrt{(\omega_a-\lambda_z)(\omega_b-2u)}/(1+\beta) $.
   We expect all the interesting phenomena in the $ U(1) $ and QT regime in Fig.\ref{crossover} at a finite $ N=3-9 $ qubits,
   especially the dramatic Berry phase effects  in both regimes can be observed in near future experiments.


\section{ Conclusions and discussions }

  Quantum optics differs from condensed matter physics at least in two important ways (1) the former mainly deal with finite size systems,
  while the latter mainly deal with thermodynamic limit ( or edge states if there is a bulk topological order )
  (2) the former mainly study pumping-decay non-equilibrium systems, while the latter mainly equilibrium systems.
  In this paper, we focused on the first feature. The combination of both features will be presented elsewhere \cite{un}.
  In studying the latter, one stress " More is different " as advocated by P. W. Anderson.
  Here, to study the four standard quantum optics models in the former system,
  we take the " Few is tricky " dual point of view \cite{berryphase,gold,comment} which establish the connections between the many body physics
  in condensed matter systems and few body problems in quantum optical systems.
  We introduced the generic $ U(1)/Z_2$ Dicke model \cite{gold} which incorporates all the 4 quantum optics models as its various special limits.
  In this paper, we investigated the new phenomena in this model at a finite $ N $  from the $ 1/J $ expansion which is complementary and dual to
  the strong coupling expansion used in \cite{strongED}.

    It is constructive to compare the two analytic methods. The instanton method employed in this paper which is in the spirit of path integral
    can map out the physical picture clearly. It starts from the $ U(1) $ limit with $ \beta=0 $ and animate
    the consecutive formation of the bound  states, quantum tunneling processes subject to the Berry phase effects shown in Fig.\ref{crossover},\ref{bound},\ref{instanton},  the energy level evolution from the $ U(1) $ to the QT regime in Fig.\ref{levelevolution}.
    It is the Berry phase interference effects which lead to the infinite oscillations in the parity of the ground state doublets in Eqn.\ref{splitting}
    and also excited state doublets Eqn.\ref{splittingn}. It can be used to predict the zeros happen
    at $ \alpha=\pm 1/2 $ and the maximum splittings happen at $ \alpha=0 $  phenomenologically, but can not used to predict where
    the zeros and maximum splittings happen in $ g/g_c $. For example, it is hard to determine the behaviors of these zeros
    as $ \beta \rightarrow 1^{-} $ limit. Only when taking the results achieved from the strong coupling expansion from the $ \beta=1 Z_2 $ limit,
    one can see that all the zeros are pushed into infinity in the $ Z_2 $ limit.
    However, we need to evaluate the photon correlations functions Eqn.\ref{aacorrtun}, \ref{aacorrtuna} and \ref{nn1}
    separately in the $ U(1) $ regime ( Fig.\ref{photonsplitting}a ) by the perturbation theory
    and in the QT regime ( Fig.\ref{photonsplitting}b ) in an intuitive and phenomenological way.
    The strong coupling expansion employed in \cite{strongED} which is in the spirit of canonical quantization
    can not distinguish the differences between
    the scattering states and the bound states, therefore not the physical process
    of the bound state formation in Fig.\ref{crossover},\ref{bound},\ref{instanton}.
    It starts from the $ Z_2 $ limit with $ \beta=1 $.
    The Berry phase effects are only implicitly embedded in the expansion in term of the anisotropic parameter
    away from the $ Z_2 $ limit $ \beta \neq 1 $. So the physical picture is less clear.
    However, it can be used to evaluate the first $ N $ zeros very precisely when compared with the ED in Fig.\ref{leveledn2} and Fig.\ref{zeros159}.
    It can also be used to calculate all the photon correlation functions in both the QT and $ U(1) $ regimes systematically  and
    in a unified scheme.
    So the two analytical methods are complementary and dual to each other. Their combination leads to rather complete understandings of
    both physical mechanisms and quantitative values of all the experimental measurable quantities in the QT regimes and the $ U(1) $ regime
    not too close to the QCP at $ N=\infty $ in Fig.\ref{crossover} and \ref{levelevolution}.

At the $ U(1) $ limit $ \beta=0 $, there are infinite level crossings due to the Berry phase effects at a finite $ N $  ( Fig.\ref{levelevolution}a ) as presented in \cite{berryphase,gold,comment}.
   Turning on  a small $ \beta $ will only lead to level repulsions between the same parity states,
   the Berry phase still leads to the level crossings between the even and odd state, therefore the alternating parities on the ground state and also all
   the doublets at $ |m|= 1, \cdots P $ in the $ U(1) $ regime ( Fig.\ref{levelevolution}b ).
   When $ \beta $ gets bigger, the system evolves into the QT regime where the Berry phase
   continue to play a crucial role leading to interference between different instanton tunneling events ( Fig.\ref{levelevolution}c and d ).
   As $ \beta $ gets close to 1, the $ U(1) $ regimes disappears, the normal state directly gets to the QT regime, the Berry phase effects show up after
   the formations of all the bound states.
   At the $ Z_2 $ limit $ \beta =1 $, $ \lambda=0 $, the Berry phase effects are pushed into infinity, so
   there is no level crossings between opposite parities anymore in Fig.\ref{leveledn2}d,
   the energy levels statistics at a given parity sector satisfy the Wigner-Dyson distribution in the superradiant regime \cite{chaos}.
   However, at any $ \beta < 1 $ in Eqn.\ref{u1z2u1}, as shown in \cite{strongED}, it is the extra term  $ \lambda ( a^{\dagger}- a ) i J_y $ which introduces frustrations, therefore Berry phase effects into the $ Z_2/U(1) $ Dicke model. They leads to infinite level crossings with alternating even and odd parity in the ground state and all the doublets at $ l > 1 $ ( Fig.\ref{levelevolution}d ).
   Combining the physical picture from $ \beta $ small achieved from $ U(1) $ limit by instanton method in this paper to
   large $ \beta \sim 1 $ achieved from $ Z_2 $ limit by the strong coupling expansion method in \cite{strongED},
   we conclude that it is the Berry phase effects which lead to the level crossings at any $ 0 \leq \beta < 1 $ except at the $ Z_2 $ limit $ \beta=1 $.
   From Fig.\ref{levelevolution}, we expect that the level statistics at a given parity sector still satisfies the Possion statistics
   in the normal regime, the Wigner-Dyson distribution  in the QT regime, but it remains  interesting to see how it changes in the $ U(1) $ regime
   in Fig.\ref{crossover}.
    It was shown in Ref.\cite{vbs} that it is Berry phase effects in the instanton tunneling events
    in the $ 2 +1 $ compact QED which leads to the Valence bond order in 2d quantum Anti-ferromagnet.
    Here we showed that it is Berry phase effects in the $ 0+1 $ dimensional instanton tunneling events in the compact phase of photons which leads to
    the infinite level crossing with alternating parity in the ground and low energy excited states ( Fig.\ref{levelevolution},\ref{leveledn2} ).


   There are some illuminating duality in both the Hamiltonian and the quantum numbers characterizing the energy spectrum
   of the $ U(1)/Z_2 $ Dicke model. In the present paper, we start from the Hamiltonian in its $ U(1)/Z_2 $ representation Eqn.\ref{u1z2u1} and
   use its complete eigenstates $ | l \rangle_m | m \rangle $ when $ \beta $ is not too large ( namely near the $ U(1) $ limit ).
   In the $ U(1) $ crossover regime in Fig.\ref{crossover} and \ref{levelevolution},
   the Landau level index $ l=0,1,...,N $ ( $ N+1 $ Landau levels ) denotes the high energy
   Higgs type of excitation, the magnetic number $ m=-P,-P+1,.... $ ( no upper bounds )  denotes the low energy pseudo-Goldstone mode \cite{gold}.
   However, in the strong coupling expansion used in \cite{strongED}, we start from the Hamiltonian in its dual $ Z_2/U(1) $ representation
   and use its complete eigenstates $ | l \rangle_m |j, m \rangle $ when $ 1-\beta $ is not too large ( namely near the $ Z_2 $ limit ).
   In the QT regime in Fig.\ref{crossover} and \ref{levelevolution}, the Landau level index $ l=0,1,...$ ( no upper bounds ) denotes the low energy
   atomic excitation, the magnetic number $ m=-j,-j+1,....,j $ ( also $ 2j+1=N+1 $ )  denotes the high energy optical mode.
   So the Landau level index and the magnetic number exchanges their roles from the $ U(1) $ to the  QT  regime.
   The crossover between the two basis is precisely described in Fig.\ref{levelevolution}.
   Of course, when $ \beta $ gets too close to 1, the $ U(1) $ regime disappears and  so does the duality relations.

{\bf Acknowledgements}

J. Ye thank Yu Chen for his participation and Prof. Guangshan Tian for encouragements in the very early stage of the project.
We thank Han Pu and Lin Tian for helpful discussions.
Y.Y and JY are supported by NSF-DMR-1161497, NSFC-11174210.
W.M. Liu is supported by NSFC under Grants No. 10934010 and No. 60978019, the NKBRSFC under Grants No. 2012CB821300.
CLZ's work has been supported by National Keystone Basic Research Program (973 Program) under Grant No. 2007CB310408,
No. 2006CB302901 and by the Funding Project for Academic
Human Resources Development in Institutions of Higher Learning Under the Jurisdiction of Beijing Municipality.







\begin{thebibliography}{99}
\bibitem{extension} Note that For the general $ U(1)/Z_2 $ Dicke model Eqn.\ref{u1z2u1} with $ 0 < \beta < 1 $,
the normal to the superradiant transitions at $ N=\infty $ share the same universality class as the $ Z_2 $ limit at $ \beta=1 $,
so only the coefficient $ c $ depends on $ \beta $.
As shown here, the dramatic qualitative differences due to $ \beta \neq 1 $
show up only away from the QCP in the $ U(1) $ and QT regime in Fig.\ref{crossover}.
It is the purpose of this manuscript to explore the new phenomena in the $ U(1) $ and QT regime at a general $ 0< \beta < 1 $
at a finite $ N $.
\end{thebibliography}
\end{document}